\documentclass[twocolumn,english,aps,pra,eqsecnum]{revtex4-2}
\usepackage[T1]{fontenc}
\usepackage[latin9]{inputenc}
\usepackage{bm}
\usepackage{amsmath}
\usepackage{amssymb}
\usepackage{graphicx}
\usepackage{geometry}
\makeatletter

\providecommand{\tabularnewline}{\\}

\makeatother

\usepackage{babel}
\begin{document}
\title{Matrix phase-space representations for quantum symmetries }
\author{Peter D. Drummond, Alexander S. Dellios, and Margaret D. Reid}
\affiliation{Centre for Quantum Science and Technology Theory, Swinburne University
of Technology, Melbourne 3122, Australia}
\email{peterddrummond@protonmail.com}

\begin{abstract}
We introduce a general phase-space representation that includes global
quantum symmetries in the basis expansion. This method, called matrix
phase-space, projects the basis onto a reduced Hilbert space, which
can greatly reduce sampling errors of many-body quantum simulations
and unifies several previous phase-space methods. The purpose of this
paper is to provide detailed proofs of basic theorems and operator
identities. We also treat several different types of symmetries. To
illustrate the benefits of matrix phase-space methods, we give a detailed
derivation of a recent application to the topical problem of verifying
the outputs of Gaussian boson sampling (GBS) quantum computers with
photon number resolving detectors. This has exponential complexity,
and using parity symmetry reduces sampling errors by very large factors
relative to earlier methods.
\end{abstract}
\maketitle

\section{Introduction}

One of the main theoretical challenges in physics is the calculation
of large-scale many-body quantum dynamics. Exact solutions are rare,
while orthogonal expansions for dynamics are not useful due to the
exponential growth of the Hilbert space dimension. Even if analytic
solutions exist, they often involve functions like permanents \citep{Aaronson2011},
Hafnians \citep{Hamilton2017gaussian} or Bethe ansatz wave-functions
\citep{yurovsky2017dissociation}. Calculating observables using such
functions is exponentially complex. They cannot be computed at large
scale.

Phase-space methods are a common solution for simulating many-body
quantum systems with large Hilbert spaces. Such methods utilize probabilistic
sampling, where operator expectation values are replaced by statistical
estimates of observables. This reduces computational complexity at
the expense of introducing sampling errors. The first complete, positive
phase-space probability distribution was the Husimi Q-distribution
\citep{Husimi1940}. Although a non-positive-definite dynamical diffusion
\citep{Altland_PRL2012_Qchaos} limits its use for dynamics, it is
important in foundational research \citep{drummond2020retrocausal}.
There are other phase-space distributions, such as the Wigner \citep{Wigner_1932}
and Glauber \citep{Glauber_1963_P-Rep}, distributions, but they are
not positive definite, and can only be efficiently sampled in some
cases.

The normally ordered positive-P (+P) representation \citep{drummond1980generalised}
has a positive distribution and diffusion term, and is used to simulate
dynamical open quantum systems \citep{drummond2016quantum}. Cases
of experimental relevance include higher dimensional and multimode
problems such as quantum solitons \citep{Carter:1987,Carter:1991},
parametric amplifiers \citep{raymer1991limits}, Bose-Einstein condensate
collisions \citep{Deuar:2007_BECCollisions}, and Gaussian boson sampling
(GBS) \citep{drummondSimulatingComplexNetworks2022}. However, there
is a trade-off here. The non-uniqueness of the basis expansion causes
sampling issues at large nonlinearity. In such cases the distribution
can develop large tails which are hard to sample efficiently, and
may cause systematic errors \citep{Gilchrist_Gardiner_PD_PPR_Application_Validity,Assaad:2005,Deuar2006a}.
This issue can even occur for high-order correlations in linear cases
with large numbers of modes \citep{drummond2026matrix}. 

A number of methods are known for solving this problem. One approach,
the stochastic Bloch representation, uses projections onto an eigenvalue
of a global number operator \citep{Carusotto:2001}, which improves
results in nonlinear Bose-Einstein quantum dynamics. Similar methods
also produce large improvements with fermionic ground-state calculations
\citep{corboz2008systematic}. Another technique is the stochastic
gauge P-representation \citep{Deuar2006a,Deuar:2002}. This representation
weights stochastic trajectories. The resulting removal of extended
tails and boundary terms can reduce sampling errors by many orders
of magnitude.

In this paper, we define a normally ordered phase-space method that
unifies group symmetry with phase-space by introducing global symmetry
projection operators. We call this the matrix P-representation. It
is a complete expansion, allowing for multiple conservation laws and
symmetries. It includes as a subset both the gauge-P and stochastic
Bloch methods. The new feature is a matrix of weights that define
the relative probability of symmetry eigenvalues.

A general formalism for both fermions and bosons is defined, and relevant
theorems and identities are proven before focusing on bosonic representations.
A crucial issue is obtaining projection operators compatible with
phase-space mappings. Using discrete Fourier transforms of quantum
states generated by quantum superpositions, we show that for bosonic
cases a mapping from cyclic groups of coherent states to a matrix
group allows one to obtain phase-space equations that are equivalent
to quantum dynamics, and can be sampled.

Including a stochastic matrix for global symmetry eigenvalues is different
to other ways of embedding a symmetry group in a phase-space. In earlier
methods, $SU(2)$ groups have been applied to large systems with probabilistic
sampling \citep{Haake:1979,DrummondEberly1981PhysRevA.25.3446,Barry_PD_qubit_SU},
but these usually involve approximate mappings. The discrete Wigner
function \citep{Antonopoulos2025} and related methods \citep{brif1998general,Brif_Mann_1999_PhaseSpaceForm_LieG},
are not positive, and are typically used for small Hilbert spaces.
In the present theory, phase-space is used for local fluctuations.
Global symmetry eigenvalues are encoded into a stochastic density
matrix, which can provide large improvements in sampling error compared
to older methods using phase-space expansions \citep{drummond2026matrix}.

Earlier work \citep{drummond2026matrix} applied matrix phase-space
to the verification of low-loss Gaussian boson sampling quantum computers
\citep{Hamilton2017gaussian}. Such devices are a potential route
towards demonstrating quantum computational advantage. A number of
experiments now claim this \citep{zhongPhaseProgrammableGaussianBoson2021,madsenQuantumComputationalAdvantage2022,deng2023gaussian,liu2025robust},
although the hardware has errors whose effect is significant \citep{drummondSimulatingComplexNetworks2022,zlokapa2023boundaries}.
Because of errors, the development of verification tools in parallel
with hardware improvements is essential, since practical error-correction
is difficult to implement \citep{google2025quantum}. More fundamentally,
if quantum mechanics requires changes at macroscopic scales \citep{ghirardi1986unified,Ghirardi_PRA1990,diosi1987universal,penrose1996gravity,marshall2003towards,diosi1989models},
quantum computer outputs may not agree with current theory, again
requiring verification.

Even though +P methods are highly efficient for validating large-scale
GBS quantum computers with losses \citep{dellios2021,dellios2025validationPLA,dellios2025validationPNR,drummondSimulatingComplexNetworks2022},
the presence of distribution tails in low-loss, number-resolved cases
causes results to become inaccurate \citep{drummond2026matrix}. Other
common phase-space representations such as the Wigner and Q-function
methods have exponentially growing sampling errors \citep{drummondSimulatingComplexNetworks2022}.
In this paper, a more detailed description of the matrix-P theory
applied to GBS is given, illustrating the advantages of using symmetries
in phase-space. More general results are also proved, including for
different symmetries and conservations, that apply to other cases
as well. 

The rest of the paper is structured as follows. In Section \ref{sec:Matrix-phase-space},
we define the general matrix representation as a positive phase-space
expansion that includes global symmetries. We then treat normally
ordered bosonic cases, which we call matrix P-representations. In
Section \ref{sec:Projected-bosonic-coherent}, we introduce projected
bases for bosons and obtain their properties, using matrix group superpositions
of coherent states. Section \ref{sec:Dynamics-and-operator} obtains
general differential identities for time evolution, and derives properties
of the projection operators that generate the basis. Section \ref{sec:Applications}
applies the method to GBS, with squeezed state inputs to a network
as a parity example. Finally, Section \ref{sec:Conclusions} gives
a summary of the paper. 

\section{matrix phase-space\label{sec:Matrix-phase-space}}

We first introduce matrix phase-space representations. These combine
a global stochastic density matrix with phase-space variables that
describe local fluctuations. Since there are only a finite number
of global symmetries, the expansion is scalable. 

Suppose a quantum system has one or multiple symmetries. If the symmetry
is exact, it leads to conservation laws via Noether's theorem \citep{Noether1918}.
Possible conservation laws include particle number and charge, arising
from a global phase symmetry, or momentum arising from a translation
symmetry. 

The expansion is not limited to exact symmetries. If the symmetry
is only partly conserved, it can still be treated as long as a group
generator exists. These methods can treat either symmetry preserving
or breaking evolution, although with reduced utility if strongly broken. 

\subsection{General definitions }

A standard phase-space representation expands a quantum many-body
density matrix as
\begin{equation}
\hat{\rho}\left(t\right)=\int P(\vec{\alpha},t)\hat{\Lambda}(\vec{\alpha})\text{d}\vec{\alpha}.\label{eq:General_phase_space}
\end{equation}

Here, $P(\vec{\alpha},t)$ is time-dependent probability density defined
on a real or complex phase-space $\vec{\alpha}$ with volume measure
$\text{d}\vec{\alpha}$, and $\hat{\Lambda}(\vec{\alpha})$ is the
kernel operator basis expanded as 
\begin{equation}
\hat{\Lambda}(\vec{\alpha})=\hat{\Lambda}^{(u)}(\vec{\alpha})e^{-w(\vec{\alpha})},
\end{equation}
where $\hat{\Lambda}^{(u)}(\vec{\alpha})$ is an unnormalized, unprojected
kernel operator and $w(\vec{\alpha})$ is a weight function, generally
defined so that the probability is normalized. 

\subsubsection{Matrix phase-space representations}

To obtain a matrix phase-space representation with $\mathcal{M}$
quantum numbers, the many-body density matrix is expanded instead
as 
\begin{equation}
\hat{\rho}(t)=\int P_{\mathcal{M}}(\vec{\lambda},t)\hat{\Lambda}_{\mathcal{M}}(\vec{\lambda})\text{d}\vec{\lambda},\label{eq:Matrix_representation}
\end{equation}
where $P_{\mathcal{M}}(\vec{\lambda},t)$ are probabilities of the
kernel operator basis 
\begin{equation}
\hat{\Lambda}_{\mathcal{M}}(\vec{\lambda})=\text{Tr}_{\mathcal{M}}\left[\underline{\Omega}\hat{\underline{\Lambda}}(\vec{\alpha})\right],\label{eq:projected basis}
\end{equation}
and $\text{d}\vec{\lambda}\equiv\text{d}\left[\vec{\alpha},\underline{\Omega}\right]$
is a volume measure over the combined phase-space, including the $\mathcal{M}\times\mathcal{M}$
complex symmetry matrix $\underline{\Omega}$. The $\mathcal{M}\times\mathcal{M}$
kernel operator matrix $\hat{\underline{\Lambda}}(\vec{\alpha})$,
is defined relative to the symmetry projectors $\hat{\Pi}_{\tilde{p}}$
and new weights $w_{\tilde{p}\tilde{q}}$ as
\begin{equation}
\hat{\Lambda}_{\tilde{p}\tilde{q}}\left(\vec{\alpha}\right)=\hat{\Pi}_{\tilde{p}}\hat{\Lambda}^{(u)}\left(\vec{\alpha}\right)\hat{\Pi}_{\tilde{q}}e^{-w_{\tilde{p}\tilde{q}}(\vec{\alpha})}.\label{eq:coherent_projection_matrix}
\end{equation}

\subsubsection{Notation}

Before proceeding, we clarify the notation. Objects in Hilbert space
follow standard quantum mechanical definitions, with operators labelled
$\hat{O}$, and expectation values denoted $\left\langle \hat{O}\right\rangle =\text{Tr}\left(\hat{\rho}\hat{O}\right)$. 

Vectors and matrices of dimension $M$ (for $M$ modes) are in bold
typeface, where latin script $i,j$ is used for individual mode labels.
The modes could include spatial positions, momenta, and spin or polarization
internal degrees of freedom. 

In quantum phase-space, $M$-dimensional coherent amplitude vectors
are denoted $\bm{\alpha}$, and we also use $\vec{\alpha}=(\boldsymbol{\alpha},\boldsymbol{\beta})$
for the double-dimensional spaces used in +P expansions. The notation
for matrix-P phase-space is $\vec{\lambda}$, which includes a stochastic
density matrix. The indices of $\boldsymbol{\alpha}$ are typically
spatial lattice positions or their Fourier transforms $(k_{x},k_{y},k_{z})$,
and spin or polarization eigenvalues, $s$, which we group into a
unified index.

The number of global symmetries is $\mathcal{N}$. Greek letters $\mu,\nu$
label different symmetry groups. The number of group elements of the
$\nu$-th group is $\mathcal{M}_{\nu}$, and $p_{\nu}=0,1,\dots,\mathcal{M}_{\nu}-1$
are their indices or quantum numbers. Since multiple symmetries can
be treated, we use quantum number vectors to symbolize these. An $\mathcal{N}$-vector
is indicated by a tilde, so that $\mathcal{\tilde{M}}\equiv[\mathcal{M}_{1},\mathcal{M}_{2},\dots,\mathcal{M}_{\mathcal{N}}]$,
and $\tilde{p}=[p_{1},p_{2},\dots,p_{\mathcal{N}}]$. Sums or products
over $\tilde{p}$ include all possible values of $\tilde{p}$.

For clarity, $\mathcal{M}\times\mathcal{M}$ matrices are underlined,
where $\mathcal{M}\equiv\prod\mathcal{M}_{\nu}$ is the total number
of quantum number combinations available, e.g. the identity matrix
$\underline{I}$. The trace of $\mathcal{M}\times\mathcal{M}$ matrices
are labelled $\text{Tr}_{\mathcal{M}}$. Trace cyclic properties are
restricted to each trace type. That is, only $\mathcal{M}\times\mathcal{M}$
matrices satisfy the cyclic property in $\text{Tr}_{\mathcal{M}}$,
while for the Hilbert space trace, $\text{Tr}\left(\dots\right)$,
the cyclic property is limited to operators and quantum density matrices.

To help the reader navigate quantum states, phase-spaces and symmetries,
we tabulate the notation used in Table (\ref{tabNotation table}).

\begin{table}
\centering{}%
\begin{tabular}{|c|c|c|}
\hline 
Linear space & Dimension & Symbol\tabularnewline
\hline 
\hline 
Normalized quantum states & $\infty$ & $\left|\psi\right\rangle $\tabularnewline
\hline 
Un-normalized coherent states & $\infty$ & $\left\Vert \bm{\alpha}\right\rangle $\tabularnewline
\hline 
Vector over modes & $M$ & $\bm{\alpha}$\tabularnewline
\hline 
Vector in matrix phase-space & $2M+\mathcal{M}^{2}$ & $\vec{\lambda}$\tabularnewline
\hline 
Vector of quantum numbers  & $\mathcal{\mathcal{N}}$ & $\tilde{p}$\tabularnewline
\hline 
Vector with quantum number index & $\mathcal{M}$ & $\underline{\left\Vert \boldsymbol{\alpha}\right\rangle }$\tabularnewline
\hline 
\end{tabular}\caption{Comparison table of notation for linear spaces.\label{tabNotation table}}
\end{table}

\subsubsection{Symmetry projections}

Next, we introduce general sets of commuting projectors $\hat{\Pi}_{\nu}$,
which encode global symmetries into the density matrix. These may
be continuous symmetries, but we will treat them as discrete, since
a continuous symmetry has discrete values as subsets. The continuous
case is obtained in the limit of $\mathcal{M}_{\nu}\rightarrow\infty.$
Discrete subsets have useful properties without taking limits, as
we show below. Non-abelian groups occur in particle physics \citep{ishimori2010non},
but are not treated here for simplicity.

As is the case for any projection operator, the $\nu$-th symmetry
projector satisfies the completeness relation
\begin{align}
\sum_{p_{\nu}=0}^{\mathcal{M}_{\nu}-1}\hat{\Pi}_{\nu,p_{\nu}} & =\hat{1},\label{eq:projector completeness-1}
\end{align}
and orthogonality relation 
\begin{equation}
\hat{\Pi}_{\nu,p_{\nu}}\hat{\Pi}_{\nu,q_{\nu}}=\delta_{p_{\nu}q_{\nu}}\hat{\Pi}_{\nu,p_{\nu}}.\label{eq:projector orthonormality}
\end{equation}

Since one can project multiple global symmetries, we define a compact
notation for the overall projector and number of eigenvalues:
\begin{align}
\hat{\Pi}_{\tilde{p}} & \equiv\prod_{\nu=1}^{\mathcal{N}}\hat{\Pi}_{\nu,p_{\nu}}\label{eq:projector_composite}\\
\mathcal{M} & \equiv\prod_{\nu=1}^{\mathcal{N}}\mathcal{M}_{\nu}.
\end{align}

The Hamiltonians and dynamical master equations do not have to satisfy
the corresponding symmetries, because the representations are complete.
For example, one can have losses, which violates parity conservation,
and we show below how this can be treated even within a parity symmetry
projection. Nevertheless, projected representations are most useful
if the quantum numbers are conserved, or if the non-conserving behavior
is small.

In Eq (\ref{eq:coherent_projection_matrix}), the bare kernel $\hat{\Lambda}^{(u)}$
can correspond to different types of Hilbert space. For example, it
could be be an unnormalized coherent projector for coherent bosonic
states \citep{Bargmann:1961,drummond1980generalised}, $SU^{(n)}$
coherent states \citep{arecchi1972atomic}, or gaussian states \citep{corney2004gaussian},
which include bosonic and fermionic cases.

\subsection{General weights}

Matrix phase-space representations are defined by three choices:
\begin{itemize}
\item The distribution $P$ and kernel $\hat{\Lambda}^{(u)}$ that defines
the base representation.
\item The symmetry projectors $\hat{\bm{\Pi}}_{\nu}$ that define the symmetry
groups.
\item The weight functions $w_{\tilde{p}\tilde{q}}\left(\vec{\alpha}\right)$
that give the relative weights of trajectories.
\end{itemize}
The weight functions are used to optimize later computations, and
a variety of weights are possible. 

\subsubsection{Normal weight}

One of the weight functions treated in this paper is the normal weight
\begin{equation}
w_{\tilde{p}\tilde{q}}^{(N)}\left(\vec{\alpha}\right)=\ln\left(\sqrt{G_{\tilde{p}}^{(N)}\left(\vec{\alpha}\right)G_{\tilde{q}}^{(N)}\left(\vec{\alpha}\right)}\right),
\end{equation}
where 
\begin{align}
G_{\tilde{p}}^{(N)}(\vec{\alpha}) & =\text{Tr}\left(\hat{\Pi}_{\tilde{p}}\hat{\Lambda}^{(u)}(\vec{\alpha})\right),\label{eq:general canonical weight}
\end{align}
denotes an inner product of unnormalized projected states. 

For normally ordered bosonic representations, this weight choice causes
matrix phase-space methods to reduce to the +P \citet{drummond1980generalised}
or gauge-P \citet{Deuar_RPA2002} representations in the case of $\mathcal{M}=1$. 

\subsubsection{Regularized weights}

If $G_{\tilde{p}}^{(N)}=0$ the normal weight can lead to singularities.
An alternative example choice that is close to the normal weight everywhere
except at the zero location is 
\begin{equation}
\frac{1}{G_{\tilde{p}}}=\frac{G_{\tilde{p}}^{(N)*}}{\left(\epsilon+\left|G_{\tilde{p}}^{(N)}\right|^{2}\right)},
\end{equation}
where $\epsilon$ is a regularizing parameter. Substituting into the
normal weight above gives 
\begin{equation}
w_{\tilde{p}\tilde{q}}^{(R)}\left(\vec{\alpha}\right)=\ln\left(\sqrt{\frac{\left(\epsilon+\left|G_{\tilde{p}}^{(N)}\right|^{2}\right)\left(\epsilon+\left|G_{\tilde{q}}^{(N)}\right|^{2}\right)}{G_{\tilde{p}}^{(N)*}G_{\tilde{q}}^{(N)*}}}\right).
\end{equation}

With this regularization, the normalizing term is finite even at zeros
of the inner product $G_{\tilde{p}}^{(N)}$, but the operator identities
are more complex. Other methods for weights are possible, and can
give lower sampling errors, depending on the physical problem \citep{Gilchrist_Gardiner_PD_PPR_Application_Validity,Deuar2006b,Deuar:2001,Deuar:2002}.
One is the ``simple'' weight of the stochastic Bloch method \citep{Carusotto:2001,carusotto2001exact,Carusotto2003,polyakov2015quasiprobability,Zin2018,carusotto2003exact},
described in greater detail below.

\subsection{Normal weight properties}

The following results hold for any matrix representation using normal
weights, and are related to normalization properties. For simplicity,
we set $w_{\tilde{p}\tilde{q}}=w_{\tilde{p}\tilde{q}}^{(N)}$ in this
section, unless otherwise stated. 

\subsubsection{Proposition (1): Kernel trace identity}

The operator trace of the kernel matrix operator $\hat{\underline{\Lambda}}$
is the $\mathcal{M}\times\mathcal{M}$ identity matrix $\underline{I}$:
\begin{equation}
\text{Tr}\left[\hat{\underline{\Lambda}}\left(\vec{\alpha}\right)\right]=\underline{I}.\label{eq:kernel trace result}
\end{equation}

\textbf{Proof:} This result follows directly by substitution. Dropping
phase-space arguments for brevity, and using cyclic ordering trace
properties,
\begin{equation}
\text{Tr}\left[\hat{\Lambda}_{\tilde{p}\tilde{q}}\right]=\text{Tr}\left[\hat{\Pi}_{\tilde{q}}\hat{\Pi}_{\tilde{p}}\hat{\Lambda}^{(u)}\left(\vec{\alpha}\right)e^{-w_{\tilde{p}\tilde{q}}}\right].
\end{equation}
Next, using Eq.(\ref{eq:projector orthonormality}) we see that:
\begin{equation}
\text{Tr}\left[\hat{\Lambda}_{\tilde{p}\tilde{q}}\right]=\delta_{\tilde{p}\tilde{q}}\text{Tr}\left[\hat{\Pi}_{\tilde{p}}\hat{\Lambda}^{(u)}\left(\vec{\alpha}\right)\hat{\Pi}_{\tilde{p}}\right]e^{-w_{\tilde{p}\tilde{p}}}.
\end{equation}
For the normal weight, $\text{Tr}\left[\hat{\Pi}_{\tilde{p}}\hat{\Lambda}^{(u)}\left(\vec{\alpha}\right)\hat{\Pi}_{\tilde{p}}\right]=e^{w_{\tilde{p}\tilde{p}}}$,
which proves the required result, since:
\begin{equation}
Tr\left[\hat{\Lambda}_{\tilde{p}\tilde{q}}\right]=\delta_{\tilde{p}\tilde{q}}e^{w_{\tilde{p}\tilde{p}}}e^{-w_{\tilde{p}\tilde{p}}}=\delta_{\tilde{p}\tilde{q}}.
\end{equation}

\subsubsection{Proposition (2): Symmetry matrix trace identity}

The symmetry matrix $\underline{\Omega}$ has unit average matrix
trace:
\begin{equation}
\text{Tr}_{\mathcal{M}}\left[\left\langle \underline{\Omega}\right\rangle _{\mathcal{M}}\right]=1,\label{eq:Average trace property}
\end{equation}
where we define $\left\langle f\right\rangle _{\mathcal{M}}\equiv\int P_{\mathcal{M}}(\vec{\lambda},t)f(\vec{\lambda})\text{d}\vec{\lambda}$,
which is the average value of $f$ relative to the given matrix phase-space
probability, $P_{\mathcal{M}}$. This result implies that in this
case the averaged symmetry matrix has the properties of a reduced
density matrix for the symmetry eigenvalues.

\textbf{Proof:} The result follows from the properties of the density
matrix operator trace, namely $\text{Tr}\left[\hat{\rho}\right]=1$.
Combined with the kernel trace result, Eq.(\ref{eq:kernel trace result})
and linearity, we see that
\begin{equation}
\text{Tr}_{\mathcal{M}}\left[\left\langle \underline{\Omega}\right\rangle _{\mathcal{M}}\right]=\int P_{\mathcal{M}}(\vec{\lambda},t)\text{Tr}_{\mathcal{M}}\left[\underline{\Omega}\right]\text{d}\vec{\lambda}=1,
\end{equation}
which proves the result required.

\subsubsection{Proposition (3): Probability normalization}

Provided the matrix $\underline{\Omega}$ has unit trace and $P_{\mathcal{M}}(\vec{\lambda},t)$
is real, the probability $P_{\mathcal{M}}(\vec{\lambda},t)$ is normalized:
\begin{equation}
\int P_{\mathcal{M}}(\vec{\lambda},t)\text{d}\vec{\lambda}=1.
\end{equation}

\textbf{Proof:} This follows from the average trace property Eq.(\ref{eq:Average trace property}),
since:
\begin{equation}
\int P_{\mathcal{M}}(\vec{\lambda},t)d\vec{\lambda}=\int P_{\mathcal{M}}(\vec{\lambda},t)\text{Tr}_{\mathcal{M}}\left[\underline{\Omega}\right]\text{d}\vec{\lambda}=1,
\end{equation}
which is the result required.

If other weights are chosen, typically either the density matrix $\underline{\Omega}$
has a trace that is not unity, or else the probability is not normalized.
These weights can still be useful where the dynamics is undamped and
nonlinear. In such cases the normal weights can lead to growing tails
with movable singularities, causing large sampling errors \citep{Gilchrist_Gardiner_PD_PPR_Application_Validity}
which can be improved by choosing other weight terms \citep{Carusotto:2001,Deuar:2002}. 

In linear cases, one can choose the normal weight, the symmetry matrix
$\underline{\Omega}$ has unit matrix trace and hence $P_{\mathcal{M}}(\vec{\lambda},t)$
is normalized.

\subsection{Matrix P-representations}

In the remainder of the paper, we consider bosonic Hilbert spaces
with normal ordering, where $\hat{\bm{a}}^{\dagger}$ $\left(\hat{\bm{a}}\right)$
are bosonic creation (annihilation) operator vectors. Despite this
restriction, many results can be extended to other phase-space mappings
and Hilbert spaces, which we save for future work. 

We take the normally ordered +P distribution as the base representation.
This is both complete and positive, with $2M$ phase-space amplitudes
$\vec{\alpha}=\left(\boldsymbol{\alpha},\boldsymbol{\beta}\right)$.
The off-diagonal terms with $\boldsymbol{\alpha}\neq\boldsymbol{\beta}^{*}$
allow coherent state superpositions to be included in the expansion.
The +P distribution generalizes the Glauber-Sudarshan P-distribution
\citep{Glauber_1963_P-Rep,Sudarshan_1963_P-Rep}, resulting in a $4M$-dimensional
real volume integral on a phase-space of twice the classical dimension. 

Following from Eq. (\ref{eq:General_phase_space}), at least one +P
distribution $P_{+}(\vec{\alpha},t)$ always exists, with the density
matrix expanded as
\begin{equation}
\hat{\rho}(t)=\int P_{+}(\vec{\alpha},t)\hat{\Lambda}(\vec{\alpha})\text{d}\vec{\alpha},\label{eq:+P_representation}
\end{equation}
where $P_{+}(\vec{\alpha},t)$ is a positive distribution. Here,
\begin{equation}
\hat{\Lambda}(\vec{\alpha})=\frac{\left\Vert \boldsymbol{\alpha}\right\rangle \left\langle \boldsymbol{\beta}^{*}\right\Vert }{\left\langle \boldsymbol{\beta}^{*}\right\Vert \left.\boldsymbol{\alpha}\right\rangle }\label{eq:+P_kernel_operator}
\end{equation}
is the base kernel operator defined in terms of unnormalized Bargmann
coherent state \citep{Schrodinger_CS,Bargmann:1961,Glauber_1963_P-Rep}:
\begin{equation}
\left\Vert \boldsymbol{\alpha}\right\rangle =\exp\left(\hat{\bm{a}}^{\dagger}\cdot\boldsymbol{\alpha}\right)\left|0\right\rangle .\label{eq:coherent states}
\end{equation}

In this case the unnormalized kernel is
\begin{equation}
\hat{\Lambda}^{(u)}\left(\vec{\alpha}\right)\equiv\left\Vert \boldsymbol{\alpha}\right\rangle \left\langle \boldsymbol{\beta}^{*}\right\Vert ,
\end{equation}
while $w(\vec{\alpha})=\ln\left(\left\langle \boldsymbol{\beta}^{*}\right\Vert \left.\boldsymbol{\alpha}\right\rangle \right)$
is the weight. 

The matrix P-representation can be readily derived following Eq.(\ref{eq:Matrix_representation}).
Using the notation 
\begin{equation}
\left\Vert \boldsymbol{\alpha}\right\rangle _{\tilde{p}}=\hat{\Pi}_{\tilde{p}}\left\Vert \boldsymbol{\alpha}\right\rangle \label{eq:projected_coherent_state general}
\end{equation}
for a projected coherent state \citep{drummond2016coherent}, the
kernel matrix operator is now 
\begin{equation}
\hat{\Lambda}_{\tilde{p}\tilde{q}}\left(\vec{\alpha}\right)=\left\Vert \boldsymbol{\alpha}\right\rangle _{\tilde{p}}\left\langle \boldsymbol{\beta}^{*}\right\Vert _{\tilde{q}}e^{-w_{\tilde{p}\tilde{q}}\left(\vec{\alpha}\right)}.\label{eq:bosonic_lambda_projector}
\end{equation}
Here, we keep the weight function choice general. The normal weight
$w_{\tilde{p}\tilde{q}}(\vec{\alpha})$ corresponds to the +P weight
$w$ in the $\mathcal{M}=1$ case with no symmetry. 

Defining $\vec{\alpha}'\equiv\left[\bm{\beta},\bm{\alpha}\right]$,
we will assume a factorization for any general weight such that
\begin{equation}
e^{w_{\tilde{p}\tilde{q}}\left(\vec{\alpha}\right)}=e^{w_{\tilde{p}}\left(\vec{\alpha}\right)+w_{\tilde{q}}\left(\vec{\alpha}'\right)}=\sqrt{G_{\tilde{p}}\left(\vec{\alpha}\right)G_{\tilde{q}}\left(\vec{\alpha}'\right)}.\label{eq:weight_factorization}
\end{equation}
For a positive probability, the number of samples needed is reduced
by requiring that the kernel is real, so that one can alternatively
define:
\begin{equation}
\hat{\Lambda}_{\tilde{p}\tilde{q}}\left(\vec{\alpha}\right)=\Re\left[\left\Vert \boldsymbol{\alpha}\right\rangle _{\tilde{p}}\left\langle \boldsymbol{\beta}^{*}\right\Vert _{\tilde{q}}e^{-w_{\tilde{p}\tilde{q}}\left(\vec{\alpha}\right)}\right].\label{eq:bosonic_lambda_projector-1}
\end{equation}

Since the symmetry matrix $\underline{\Omega}$ modifies trajectory
probabilities, one can effectively include part of the weight function
in this term, giving another way to change the trajectory dynamics
\citep{Deuar2006b,Deuar:2001,Deuar:2002}, known as a stochastic gauge.

Therefore, the matrix $\underline{\Omega}$ is an extension of the
complex gauge amplitude introduced in the gauge P-representation \citep{Deuar2006a,Deuar:2002,Montina2003gaugeP,PlimakOC:2001}.
In matrix representations, one can have different gauges for each
combination of symmetry eigenvalues. If we only need diagonal elements
of $\underline{\Omega}$, which depends on the density matrix, the
dynamics and the observables, this is indicated by a single quantum
number vector, so $\Omega_{\tilde{p}}\equiv\Omega_{\tilde{p}\tilde{p}}$.

\subsubsection{Normal weight}

For the normal weight choice of Eq. (\ref{eq:general canonical weight}),
one has $G_{\tilde{p}}^{(N)}\left(\vec{\alpha}\right)=G_{\tilde{p}}^{(N)}\left(\vec{\alpha}'\right)$,
since the weight combines terms with products of $\boldsymbol{\alpha}$
and $\bm{\beta}$, and one normalizes the kernel, so that:
\begin{equation}
G_{\tilde{p}}^{(N)}\left(\vec{\alpha}\right)\equiv\left\langle \boldsymbol{\beta}^{*}\right\Vert _{\tilde{p}}\left.\boldsymbol{\alpha}\right\rangle _{\tilde{p}}.\label{eq:Inner-product-normalt}
\end{equation}

For this weight choice the symmetry matrices give a reduced density
matrix if the original equations of motion preserve probability. For
example, evolution under a master equation preserves probability,
but evolution in imaginary time \citep{DrummondDK:2004} does not. 

Without projections, i.e. $\mathcal{M}=1$, the normal weight becomes
$w_{\tilde{p}\tilde{q}}^{(N)}=\boldsymbol{n}=\boldsymbol{\beta}\cdot\boldsymbol{\alpha}$
and the matrix-P representation reduces to the gauge-P representation
\citep{Deuar:2002,Deuar2005,Deuar2006b}
\begin{equation}
\hat{\rho}(t)=\int P(\vec{\lambda},t)\Omega\hat{\Lambda}(\vec{\alpha})\text{d}\vec{\lambda},\label{eq:GaugeP_representation}
\end{equation}
where the kernel operator is defined in Eq.(\ref{eq:+P_kernel_operator}).
Additionally, when $\Omega=1$ it is clear that one obtains the +P
expansion Eq.(\ref{eq:+P_representation}). 

Since it is possible to describe any existing phase-space mapping
in the form of Eq. (\ref{eq:Matrix_representation}), Table (\ref{tab:A-comparison-table})
shows the relationship of some other approaches with the matrix-P
expansion. 

\begin{table*}[t]
\centering{}%
\begin{tabular}{|c|c|c|c|c|c|}
\hline 
Representation & Ordering & Complete? & Positive? & Stochastic? & Projected?\tabularnewline
\hline 
\hline 
Wigner \citep{Wigner_1932,HILLERY1984121} & Symmetric & Y & N & N & N\tabularnewline
\hline 
Q-function \citep{Husimi1940} & Anti-normal & Y & Y & N & N\tabularnewline
\hline 
P-function \citep{Glauber_1963_P-Rep,Sudarshan_1963_P-Rep} & Normal & N & N & N & N\tabularnewline
\hline 
Positive-P \citep{drummond1980generalised} & Normal & Y & Y & Y & N\tabularnewline
\hline 
Stochastic Bloch \citep{Carusotto:2001} & Normal & N & Y & Y & Y\tabularnewline
\hline 
Gauge-P \citep{Deuar:2002,Deuar2006b} & Normal & Y & Y & Y & N\tabularnewline
\hline 
Gaussian Bose \citep{corney2003gaussian} & Normal & Y & Y & Y & N\tabularnewline
\hline 
Gaussian Fermi \citep{corney2004gaussian} & Normal & Y & Y & Y & N\tabularnewline
\hline 
Gaussian Fermi projection \citep{assaad2005symmetry} & Normal & N & Y & Y & Y\tabularnewline
\hline 
\textbf{Matrix-P} & \textbf{Normal} & \textbf{Y} & \textbf{Y} & \textbf{Y} & \textbf{Y}\tabularnewline
\hline 
\end{tabular}\caption{Comparison table of common phase-space representations. All are bosonic
except the gaussian Fermi and gaussian projected methods. Ordering
refers to the equivalent operator ordering in Dirac's sense \citep{dirac1945analogy}.
Complete means a complete representation without singularities. Positive
means the distribution is positive and probabilistic. Stochastic means
that nonlinear dynamical equations have stochastic equivalents. Projected
means the expansion includes projections.\label{tab:A-comparison-table}}
\end{table*}

\subsubsection{Simple weight}

Another weight choice is the ``simple'' weight, where one normalizes
the two states individually, so $G_{\tilde{p}}^{(S)}\left(\vec{\alpha}\right)\neq G_{\tilde{p}}^{(S)}\left(\vec{\alpha}'\right)$,
choosing instead \citet{Carusotto:2001} the diagonal inner product:
\begin{equation}
G_{\tilde{p}}^{(S)}\left(\vec{\alpha}\right)\equiv\left\langle \boldsymbol{\alpha}\right\Vert _{\tilde{p}}\left.\boldsymbol{\alpha}\right\rangle _{\tilde{p}}.\label{eq:Inner-product-simple}
\end{equation}

This weight is positive semi-definite, but the trajectories do not
all have the same probability. As a result, the individual stochastic
matrices are not reduced density matrices. As detailed below, the
$\mathcal{M}\rightarrow\infty$ limit of a phase symmetry corresponds
to total number conservation. In this limit, the matrix-P expansion
includes the ``simple'' normalization scheme of the stochastic Bloch
representation as a subset \citep{Carusotto:2001,Carusotto2003,carusotto2003exact},
although it treats more general states than just a single number eigenstate.

\subsection{Matrix P-representation existence theorem}

To prove that a bosonic matrix P-representation always exists, we
will use the result that a +P distribution always exists \citep{drummond1980generalised}
for any density matrix. 

\subsubsection{Proposition (4): Existence theorem}

For a complete set of projections $\hat{\Pi}_{\tilde{p}}$, a projected
coherent state expansion exists for any pure state, and a positive,
semi-definite bosonic matrix P-representation exists for any quantum
density matrix, and weight $w_{\tilde{p}\tilde{q}}$ .

\textbf{Proof:} From the completeness relation Eq.(\ref{eq:projector completeness-1}),
we can expand any coherent state as:
\begin{equation}
\left\Vert \boldsymbol{\alpha}\right\rangle =\sum_{\tilde{p}}\hat{\Pi}_{\tilde{p}}\left\Vert \boldsymbol{\alpha}\right\rangle =\sum_{\tilde{p}}\left\Vert \boldsymbol{\alpha}\right\rangle _{\tilde{p}}.
\end{equation}
From the completeness properties of coherent states \citep{Glauber_1963_P-Rep},
there is a unique analytic function of $\boldsymbol{\alpha}^{*}$,
$\psi\left(\boldsymbol{\alpha}^{*}\right)=\left\langle \boldsymbol{\alpha}\right\Vert \left.\psi\right\rangle $,
such that:
\begin{equation}
\left|\psi\right\rangle =\frac{1}{\pi}\int\psi\left(\boldsymbol{\alpha}^{*}\right)e^{-\left|\boldsymbol{\alpha}\right|^{2}}\left\Vert \boldsymbol{\alpha}\right\rangle d\boldsymbol{\alpha}.
\end{equation}
One can therefore expand an arbitrary pure bosonic state $\left|\psi\right\rangle $
in projected coherent states, where:
\begin{equation}
\left|\psi\right\rangle =\frac{1}{\pi}\sum_{\tilde{p}}\int\psi\left(\boldsymbol{\alpha}^{*}\right)\left\Vert \boldsymbol{\alpha}\right\rangle _{\tilde{p}}d\boldsymbol{\alpha}.
\end{equation}

Similarly, the +P kernel of Eq.(\ref{eq:+P_kernel_operator}) can
be expanded as
\begin{align}
\hat{\Lambda}(\vec{\alpha}) & =\left(\sum_{\tilde{p}}\hat{\Pi}_{\tilde{p}}\left\Vert \boldsymbol{\alpha}\right\rangle \left\langle \boldsymbol{\beta}^{*}\right\Vert \sum_{\tilde{q}}\hat{\Pi}_{\tilde{q}}\right)e^{-n}.
\end{align}
Given a weight function $w_{\tilde{p}\tilde{q}}$, this can be rewritten
by taking a stochastic density matrix $\Omega_{\tilde{p}\tilde{q}}=e^{w_{\tilde{p}\tilde{q}}-n}$,
hence
\begin{equation}
\hat{\Lambda}(\vec{\alpha})=\sum_{\tilde{q}\tilde{p}}\left[\Omega_{\tilde{p}\tilde{q}}\left\Vert \boldsymbol{\alpha}\right\rangle _{\tilde{p}}\left\langle \boldsymbol{\beta}^{*}\right\Vert _{\tilde{q}}e^{-w_{\tilde{p}\tilde{q}}}\right].
\end{equation}
Recalling that a double summation corresponds to a trace of matrix
products, and the +P expansion in Eq.(\ref{eq:+P_representation})
always exists, this gives an example of the matrix-P expansion Eq.(\ref{eq:Matrix_representation})
provided one chooses $P_{\mathcal{M}}(\vec{\lambda})$ as the following
positive semi-definite distribution:
\begin{equation}
P_{\mathcal{M}}(\vec{\lambda})=P_{+}(\boldsymbol{\alpha},\boldsymbol{\beta})\prod_{\tilde{p}\tilde{q}}\delta\left(\Omega_{\tilde{p}\tilde{q}}-e^{w_{\tilde{p}\tilde{q}}-n}\right).\label{eq:matrixP-existence}
\end{equation}

This choice is not always the optimum choice, since the existence
theorem does not guarantee uniqueness. For example, in fermionic cases
it is known that simply projecting onto symmetries without additional
changes to the equations of motion may not completely eliminate distribution
tails \citep{corboz2008systematic,Imada_2007_GBMC}. More compact
and efficient choices are often available. As we show in the operator
identities, dynamical evolution in phase-space can depend on the symmetry
eigenvalues. As a result, the factorization of Eq (\ref{eq:matrixP-existence})
is not always valid, since more complex correlated expansions of the
density matrix also exist.

\section{Projected coherent states\label{sec:Projected-bosonic-coherent}}

Up to this point we have kept the projector definition general. This
is deliberate, as there may be multiple ways to define symmetry projectors.
Here, we treat bosonic coherent state superpositions. These are useful
as they allow us to encode discrete symmetries as Fourier transforms
via a unitary transformation. 

We first recall the notation from Table (\ref{tabNotation table}),
such that:
\begin{itemize}
\item The coherent amplitude vectors have standard mode indices.
\item For more than one symmetry group the quantum numbers are also vectors. 
\item The quantum number vectors are indices for the projected states. 
\end{itemize}
For the group theorist, some results below will be familiar, but we
give them here to make the presentation self-contained.

\subsection{Group properties}

To apply phase-space representations to dynamical calculations, it
is necessary to obtain the relevant operator identities. We start
by examining the relationship between quantum states in Hilbert space
and unitary matrix transformations on the coherent state amplitudes.

\subsubsection{Unitary symmetries}

We now define the symmetry groups used here. Consider a set of $\mathcal{N}$
unitary matrix transformations $\bm{U}_{\nu}=\exp\left(i\phi_{\nu}\bm{g}_{\nu}\right)$
\citep{tung1985group,schwichtenberg2018physics} acting on the phase-space.
For the symmetry groups in this paper, their general structure is
determined by the $M\times M$ hermitian matrix $\bm{g}_{\nu}$. Following
standard conventions \citep{tung1985group} we call these the $M$-mode
group generators, and we assume they all commute. 

For Lie groups $\phi_{\nu}$ is a continuous, real phase parameter.
Here we treat discrete groups where $\phi_{\nu}=2\pi/\mathcal{M}_{\nu}$
since discrete groups are subsets of continuous groups, meaning $\boldsymbol{g}_{\nu}$
is also the generator of the continuous symmetry of interest. 

Each $\bm{U}_{\nu}$ generates a cyclic group with finite period $\mathcal{M}_{\nu}$
such that:
\begin{align}
e^{2\pi i\bm{g}_{\nu}} & =I,\label{eq:cyclic property-1}
\end{align}
and so each group $\nu$ is isomorphic to $\mathbb{Z_{\mathcal{M}_{\nu}}}$.
From the spectral theorem there is always a unitary matrix $\boldsymbol{V}$
that simultaneously diagonalizes all of the commuting $\bm{g}_{\nu}$
matrices, giving an $\mathcal{N}$-vector $\tilde{\bm{h}}$ of diagonal
$M\times M$ matrices, such that 
\begin{equation}
\tilde{\bm{g}}=\bm{V}\tilde{\bm{h}}\bm{V}^{\dagger}.\label{eq:diagonal-h-matrix-1}
\end{equation}
On expanding the resulting power series for $\bm{U}_{\nu}$ in powers
of $\phi_{\nu}$, we immediately see that for any constant $q$,
\begin{equation}
\bm{U}_{\nu}^{q}=\bm{V}\exp\left(i\phi_{\nu}q\bm{h}_{\nu}\right)\bm{V}^{\dagger}.\label{eq:diagonal-U-1}
\end{equation}
From the cyclic property, $e^{2\pi i\bm{h}_{\nu}}=\bm{I}$, so each
diagonal eigenvalue $h_{k\nu}$ is an integer, which we call the $\nu$-th
quantum number of mode $k$. This gives the quantum numbers of a single
particle occupying that mode. We also denote this as $\tilde{h}_{k}$,
to obtain an $\mathcal{N-}$vector of all the quantum numbers of a
given particle.

\subsection{Group equivalence}

The first property we examine is the group equivalence. This states
that the unitary matrix transformations on coherent state amplitudes
is equivalent to a unitary operator acting on quantum states in Hilbert
space. While this general property is known as Schur's lemma \citep{Schur1905}
in group theory literature, here we give a direct proof to show how
symmetry groups are encoded onto coherent states. 

\subsubsection{Proposition (5): Hilbert and vector space group equivalence}

The operators $\hat{g}_{\nu}\equiv\hat{\bm{a}}^{\dagger}\bm{g}_{\nu}\hat{\bm{a}}$
are commuting hermitian operators, and for $q_{\nu}=0,\ldots\mathcal{M}_{\nu}-1$,
each finite cyclic matrix group of unitary transformations $\bm{U}_{\nu}^{q_{\nu}}=\exp\left(iq_{\nu}\phi_{\nu}\bm{g}_{\nu}\right)$
on non-vanishing vectors $\boldsymbol{\alpha}\in\mathbb{C}^{M}$ is
isomorphic to the finite cyclic group of unitary operator transformations
$\hat{U}_{\nu}^{q_{\nu}}=\exp\left(iq_{\nu}\phi_{\nu}\hat{g}_{\nu}\right)$
.

\textbf{Proof:} We first prove this in the single symmetry case, as
a similar result follows for multiple symmetries. Defining a transformed
vector of bosonic operators $\hat{\bm{b}}=\boldsymbol{V}^{\dagger}\hat{\boldsymbol{a}}$,
gives 
\begin{equation}
\hat{g}=\hat{\bm{a}}^{\dagger}\bm{g}\hat{\bm{a}}=\hat{\bm{b}}^{\dagger}\bm{h}\hat{\bm{b}}.
\end{equation}
Since $\hat{\bm{b}}$ is a vector of bosonic operators, there must
exist a transformed Bargmann coherent state 
\begin{equation}
\left\Vert \bm{\alpha}\right\rangle =\exp\left(\hat{\bm{a}}^{\dagger}\bm{\alpha}\right)\left|0\right\rangle =\exp\left(\hat{\bm{b}}^{\dagger}\bm{\gamma}\right)\left|0\right\rangle \equiv\left\Vert \bm{\gamma}\right\rangle ^{(b)},\label{eq:Gamma-alpha equivalence}
\end{equation}
where $\bm{\gamma}=\bm{V}^{\dagger}\boldsymbol{\alpha}$ is the transformed
coherent state eigenvalue and $\hat{\boldsymbol{b}}^{\dagger}=\hat{\bm{a}}^{\dagger}\boldsymbol{V}$. 

Upon defining the vector of states 
\begin{equation}
\left\Vert \bm{U}^{q}\bm{\alpha}\right\rangle =\exp\left(\hat{\bm{a}}^{\dagger}\bm{U}^{q}\bm{\alpha}\right)\left|0\right\rangle ,
\end{equation}
one can substitute the diagonalization to obtain 
\begin{align}
\left\Vert \bm{U}^{q}\bm{\alpha}\right\rangle  & =\exp\left(\hat{\bm{b}}^{\dagger}\exp\left(i\phi q\right)\bm{\gamma}\right)\left|0\right\rangle =\left\Vert \exp\left(i\phi q\right)\bm{\gamma}\right\rangle ^{(b)}.
\end{align}

This result shows that the ket $\left\Vert \bm{U}^{q}\bm{\alpha}\right\rangle $
is related to the transformed coherent state Eq.(\ref{eq:Gamma-alpha equivalence})
via a phase rotation. Therefore, using the standard result for phase
rotations of a coherent state that $\left\Vert \alpha_{i}\exp(i\phi)\right\rangle =\exp(i\phi\hat{b}_{i}^{\dagger}\hat{b}_{i})\left\Vert \alpha_{i}\right\rangle $,
and applying this to each mode,
\begin{equation}
\left\Vert \bm{U}^{q}\bm{\alpha}\right\rangle =\exp\left(i\phi q\hat{\bm{b}}^{\dagger}\bm{h}\hat{\bm{b}}\right)\left\Vert \bm{\gamma}\right\rangle ^{(b)}=\exp\left(iq\phi\hat{g}\right)\left\Vert \bm{\alpha}\right\rangle .\label{eq:translated-states}
\end{equation}

The states $\left\Vert \bm{U}^{q}\bm{\alpha}\right\rangle $ in Hilbert
space are a finite cyclic group generated by the unitary operator
$\hat{U}=\exp\left(i\phi\hat{g}\right)$, where $\hat{g}=\hat{\bm{b}}^{\dagger}\bm{h}\hat{\bm{b}}=\hat{\bm{a}}^{\dagger}\bm{g}\hat{\bm{a}}$.
From the cyclic property and the completeness of the coherent state
basis, it follows that the group generated by the operator $\hat{U}$
is cyclic, and
\begin{equation}
\hat{U}^{\mathcal{M}}=\exp\left(2\pi i\hat{g}\right)=\hat{1}.\label{eq:N-th-root-operator}
\end{equation}

This demonstrates that the group of unitary transformations on coherent
amplitudes $\bm{\alpha}$ generated by the unitary matrix $\bm{U}$
generates an equivalent group of unitary transformations $\hat{U}$
on the Hilbert space. 

For a set of commuting symmetry matrices $\bm{g}_{\nu}$, it follows
from the commutation relations for bosons that $\hat{g}_{\nu}$ is
a commuting set of operators. Hence, the equivalence between groups
holds for any composite symmetry, where the unitary matrix is:
\begin{equation}
\bm{U}^{\left(\tilde{q}\right)}=\exp\left(i\sum_{\nu=1}^{\mathcal{N}}q_{\nu}\phi_{\nu}\bm{g}_{\nu}\right)=\exp\left(i\tilde{q}\tilde{\phi}\tilde{\bm{g}}\right),
\end{equation}
using a notation such that a product of three $\mathcal{N}$-vectors
implies a sum over their indices, and the corresponding unitary operator
is:
\begin{equation}
\hat{U}^{\left(\tilde{q}\right)}=\exp\left(i\sum_{\nu=1}^{\mathcal{N}}q_{\nu}\phi_{\nu}\hat{g}_{\nu}\right)=\exp\left(i\tilde{q}\tilde{\phi}\tilde{\hat{g}}\right).
\end{equation}

\subsection{Symmetry projection operators}

Using the group equivalence property above, we can now obtain the
symmetry projection operator. For a single symmetry group, the projected
state $\left\Vert \boldsymbol{\alpha}\right\rangle _{p}$ is a superposition
of $\mathcal{M}$ coherent states, defined as a discrete Fourier transform:
\begin{equation}
\left\Vert \boldsymbol{\alpha}\right\rangle _{p}\equiv\frac{1}{\mathcal{\mathcal{M}}}\left[\sum_{q=0}^{\mathcal{\mathcal{M}}-1}e^{i\phi q\left(\hat{g}-p\right)}\right]\left\Vert \boldsymbol{\alpha}\right\rangle ,\label{eq:projected-state}
\end{equation}
where from Eq.(\ref{eq:translated-states}), $\exp\left(i\phi q\hat{g}\right)\left\Vert \boldsymbol{\alpha}\right\rangle \equiv\left\Vert \bm{U}^{q}\bm{\alpha}\right\rangle $. 

For multiple symmetries, one instead has 
\begin{equation}
\left\Vert \boldsymbol{\alpha}\right\rangle _{\tilde{p}}\equiv\frac{1}{\mathcal{\mathcal{M}}}\sum_{\tilde{q}}e^{i\tilde{\phi}\tilde{q}\left(\hat{g}-\tilde{p}\right)}\left\Vert \boldsymbol{\alpha}\right\rangle ,\label{eq:multipple symmetry projected CS}
\end{equation}
recalling our simplified notation such that 
\begin{equation}
\sum_{\tilde{q}}\equiv\sum_{q_{1}=0}^{\mathcal{\mathcal{M}}_{1}-1}\sum_{q_{2}=0}^{\mathcal{\mathcal{M}}_{2}-1}\dots\sum_{q_{\mathcal{N}}=0}^{\mathcal{\mathcal{M}}_{\mathcal{N}}-1}.
\end{equation}
For notational reasons, we also define symmetry-projected kets to
exist outside the index range of $0,\ldots\mathcal{M}-1$, in which
case the index is interpreted modulo $\mathcal{M}$, so that $\left\Vert \boldsymbol{\alpha}\right\rangle _{\mathcal{M}}=\left\Vert \boldsymbol{\alpha}\right\rangle _{0}$
and $\left\Vert \boldsymbol{\alpha}\right\rangle _{-1}=\left\Vert \boldsymbol{\alpha}\right\rangle _{\mathcal{M}-1}$
.

Since we are defining the projection as a discrete Fourier transform,
the inverse transformation recovers the original vector of unitarily
rotated coherent states:
\begin{equation}
\left\Vert \bm{U}^{(\tilde{q})}\boldsymbol{\alpha}\right\rangle \equiv\sum_{\tilde{q}}e^{i\tilde{p}\tilde{q}\tilde{\phi}}\left\Vert \boldsymbol{\alpha}\right\rangle _{\tilde{p}}.\label{eq:Inverse transform}
\end{equation}

In Appendix A, we prove that the expansion Eq.(\ref{eq:multipple symmetry projected CS})
defines projection operators of the form of Eq.(\ref{eq:projected_coherent_state general}),
which satisfy completeness and orthogonal conditions. In Appendix
B, we provide examples of projections onto phase and translation global
symmetries. This work focuses mostly on phase symmetries in terms
of the analytical and numerical results presented. Matrix-P applications
to other symmetries and physical systems we save for future work. 

\subsection{Generator eigenstate and eigenvalue properties }

The Hilbert space generators $\hat{g}$ are responsible for the symmetry
properties of the unitary operators. Therefore, it is useful to investigate
their eigenvalues and eigenstates. Similar results are known in group
theory \citep{tung1985group}. However the proofs in this subsection
focus on applications to coherent state superpositions.

\subsubsection{Proposition (6): Eigenstates of the group unitary}

The superposition state $\left\Vert \boldsymbol{\alpha}\right\rangle _{p}$
is an eigenstate of $\hat{U}$ with eigenvalue $e^{i\phi p}.$

\textbf{Proof:} From the property Eq.(\ref{eq:N-th-root-operator}),
since $p$ is an integer, then $e^{i\phi\mathcal{M}\left(\hat{g}-p\right)}=e^{2\pi i\left(\hat{g}-p\right)}=\hat{1}$.
Using this to change the limits of the summation, 
\begin{align}
\left\Vert \boldsymbol{\alpha}\right\rangle _{p} & =\frac{1}{\mathcal{\mathcal{M}}}\left[\sum_{q=1}^{\mathcal{\mathcal{M}}}e^{i\phi q\left(\hat{g}-p\right)}\right]\left\Vert \boldsymbol{\alpha}\right\rangle .
\end{align}

Comparing this result to the summation in the original projected state
definition of Eq. (\ref{eq:projected-state}), it follows that 
\begin{equation}
\left\Vert \boldsymbol{\alpha}\right\rangle _{p}=e^{i\phi\left(\hat{g}-p\right)}\left\Vert \boldsymbol{\alpha}\right\rangle _{p}.
\end{equation}
Re-arranging the equation, one obtains:
\begin{equation}
e^{i\phi\hat{g}}\left\Vert \boldsymbol{\alpha}\right\rangle _{p}=e^{i\phi p}\left\Vert \boldsymbol{\alpha}\right\rangle _{p}.\label{eq:General-eigenvalue-equation}
\end{equation}
Hence, $\left\Vert \boldsymbol{\alpha}\right\rangle _{p}$ is an eigenstate
of $\hat{U}=e^{i\phi\hat{g}}$, with eigenvalue $e^{i\phi p}$. This
can be recursively extended to multiple symmetries, by applying the
argument to each projection successively to give that:
\begin{equation}
e^{i\sum_{\nu}\phi_{\nu}\hat{g}_{\nu}}\left\Vert \boldsymbol{\alpha}\right\rangle _{\tilde{p}}=e^{i\sum_{\nu}\phi_{\nu}p_{\nu}}\left\Vert \boldsymbol{\alpha}\right\rangle _{\tilde{p}}.\label{eq:nu-th eigenvalue}
\end{equation}

\subsubsection{Proposition (7): Continuum limit eigenvalue }

In the limit of $\mathcal{M}_{\nu}\rightarrow\infty$, the state $\left\Vert \boldsymbol{\alpha}\right\rangle _{\tilde{p}}$
is an eigenstate of $\hat{g}_{\nu}$ with eigenvalue $p_{\nu}$.

\textbf{Proof:} In the large $\mathcal{M}_{\nu}$ limit, $\phi_{\nu}\rightarrow0$,
so both sides of Eq.(\ref{eq:General-eigenvalue-equation}) can be
expanded as a power series in $\phi_{\nu}$. The projected state
in this limit is therefore an eigenstate of $\hat{g}_{\nu}$: 
\begin{equation}
\lim_{\mathcal{M}_{\nu}\rightarrow\infty}\hat{g}_{\nu}\left\Vert \boldsymbol{\alpha}\right\rangle _{\tilde{p}}=p_{\nu}\left\Vert \boldsymbol{\alpha}\right\rangle _{\tilde{p}}.\label{eq:continuum limit eigenvalue}
\end{equation}
Hence, in the continuous symmetry limit, the projected states $\left\Vert \boldsymbol{\alpha}\right\rangle _{\tilde{p}}$
are eigenstates of the Hilbert space group generators. This gives
a relationship between the physical properties of the projected coherent
states and the symmetries of the Hamiltonian. For example, if $\hat{g}$
corresponds physically to the number operator or momentum operator,
then the corresponding eigenvalue $p$ is proportional to the number
or momentum eigenvalue.

\subsection{Inner products for general symmetries}

Since the projected states are unnormalized, one must normalize the
kernel matrix using weights, requiring an evaluation of inner products.
From projection operator orthogonality in Eq.(\ref{eq:projector orthonormality}),
only identical projection eigenvalues need to be considered.

We assume in this subsection that the mode basis is chosen so that
the group generators are diagonal, and recall that the two normalizations
used here from Eqs. (\ref{eq:Inner-product-normalt}) and (\ref{eq:Inner-product-simple})
are:
\begin{align}
G_{p}^{(N)}\left(\vec{\alpha}\right) & \equiv\left\langle \boldsymbol{\beta}^{*}\right\Vert _{p}\left.\boldsymbol{\alpha}\right\rangle _{p}\nonumber \\
G_{p}^{(S)}\left(\vec{\alpha}\right) & \equiv\left\langle \boldsymbol{\alpha}\right\Vert _{p}\left.\boldsymbol{\alpha}\right\rangle _{p}.
\end{align}
From this, it is clear that the results for the simple case are always
obtainable from the normal weight results if $\boldsymbol{\alpha}=\boldsymbol{\beta}^{*}$.
Hence, we focus on the normal weight case. 

The inner product of two unnormalized coherent states is: 
\begin{equation}
\left\langle \boldsymbol{\beta}^{*}\right\Vert \left.\boldsymbol{\alpha}\right\rangle =e^{\boldsymbol{n}}.
\end{equation}
Including unitary transformations of the coherent amplitudes, one
obtains 
\begin{align}
\left\langle \bm{U}^{(\tilde{q})}\boldsymbol{\beta}^{*}\right\Vert \left.\bm{U}^{(\tilde{q}')}\boldsymbol{\alpha}\right\rangle  & =\exp\left(\boldsymbol{\beta}\cdot\bm{U}^{(\tilde{q}'-\tilde{q})}\boldsymbol{\alpha}\right).
\end{align}
Next, assuming diagonal generators with a vector of diagonal eigenvalues
$\tilde{g}_{k}$ in mode $k$:
\begin{align}
\left\langle \bm{U}^{(\tilde{q})}\boldsymbol{\beta}^{*}\right\Vert \left.\bm{U}^{(\tilde{q}')}\boldsymbol{\alpha}\right\rangle  & =\exp\left(\sum_{k}\beta_{k}\cdot e^{i\left(\tilde{q}'-\tilde{q}\right)\tilde{\phi}\tilde{g}_{k}}\alpha_{k}\right),
\end{align}
so the inner product becomes
\begin{equation}
G_{\tilde{p}}^{(N)}=\frac{1}{\mathcal{\mathcal{M}}}\sum_{\tilde{q}}e^{i\tilde{p}\tilde{q}\tilde{\phi}}\exp\left(\sum_{k}\beta_{k}\cdot e^{-i\tilde{g}_{k}\tilde{q}\tilde{\phi}}\alpha_{k}\right).
\end{equation}

Defining the stochastic occupation number $n_{k}=\alpha_{k}\beta_{k}$,
we obtain a general inner product result:
\begin{align}
G_{\tilde{p}}^{(N)} & =\frac{1}{\mathcal{\mathcal{M}}}\sum_{\tilde{q}}e^{i\tilde{p}\tilde{q}\tilde{\phi}}\prod_{k}\left(\sum_{m_{k}=0}^{\infty}\frac{n_{k}^{m_{k}}}{m_{k}!}e^{-im_{k}\tilde{h}_{k}\tilde{q}\tilde{\phi}}\right).\label{eq:general-inner-product}
\end{align}

\subsection{Physical explanation of inner product results}

To understand the physical meaning of this result, consider the limit
of $\mathcal{\mathcal{\tilde{M}}}\rightarrow\infty$, giving an $\mathcal{N}$-dimensional
Fourier transform. The above result then reduces to a simpler expression,
giving a multinomial if $h_{k}=1$. More generally it is expressed
as
\begin{align}
G_{\tilde{p}}^{(N)} & =\sum_{\bm{m}=0}^{\infty}\delta_{\tilde{p}-\sum_{k}m_{k}\tilde{h}_{k}}^{\mathcal{N}}\prod_{k=1}^{M}\frac{n_{k}^{m_{k}}}{m_{k}!}.
\end{align}

Here, the $\mathcal{N}$-dimensional Kronecker delta-function, $\delta^{\mathcal{N}}$,
selects those terms in the summation that have occupation numbers
$\bm{m}$ which satisfy all $\mathcal{N}$ conservation laws. This
requires them to give a total quantum number vector of $\tilde{p}$
after summing over the $m_{k}$ bosons occupying the $k$-th mode,
each with individual mode quantum numbers of $\tilde{h}_{k}$.

This result also has another form,
\begin{align}
G_{\tilde{p}}^{(N)} & =\left[\prod_{\nu}z_{\nu}^{p_{\nu}}\right]\exp\left(\sum_{k=1}^{M}n_{k}\sum_{\nu=1}^{\mathcal{N}}z_{\nu}^{h_{k,\nu}}\right).
\end{align}
The notation $\left[z^{p}\right]$ means the $p$-th coefficient
of the power series in $z$. 

To derive operator identities for use in dynamics, treated below,
it is necessary to take derivatives with respect to phase-space variables.
For a normal weight and one symmetry this is
\begin{align}
\frac{\partial G_{p}^{(N)}}{\partial\alpha_{k}} & =\frac{\beta_{k}}{\mathcal{\mathcal{M}}}\sum_{q=0}^{\mathcal{\mathcal{M}}-1}e^{i(p-h_{k})q\phi}\prod_{k}\exp\left(n_{k}e^{-ih_{k}q\phi}\right)\nonumber \\
 & =\beta_{k}G_{p-h_{k}}^{(N)}.
\end{align}
An analogous result holds for multiple symmetries:
\begin{equation}
\frac{\partial G_{\tilde{p}}^{(N)}}{\partial\alpha_{k}}=\beta_{k}G_{\tilde{p}-\tilde{h}_{k}}^{(N)}.\label{eq:inner-product-derivative}
\end{equation}

The ``simple'' normalization is recovered by taking $\beta_{k}=\alpha_{k}^{*}$.

\subsection{Inner products with a phase symmetry\label{subsec:Inner-products-phase-sym}}

For a phase symmetry with $\tilde{\bm{g}}=\bm{I}$ (see Appendix B),
from Eq.(\ref{eq:general-inner-product}) the normal inner product
is:
\begin{align}
G_{p}^{(N)} & =\frac{1}{\mathcal{\mathcal{M}}}\sum_{q=0}^{\mathcal{\mathcal{M}}-1}\sum_{m_{k}=0}^{\infty}e^{i(p-m)q\phi}\frac{\boldsymbol{n}^{m}}{m!}.
\end{align}

Using the periodic Kronecker delta function property
\begin{equation}
\delta_{p,m}^{(\mathcal{M})}=\sum_{j=-\infty}^{\infty}\delta_{m+j\mathcal{\mathcal{M}},p}=\frac{1}{\mathcal{\mathcal{M}}}\sum_{q=0}^{\mathcal{\mathcal{M}}-1}e^{i\left(p-m\right)q\phi},\label{eq:periodic_kronecker}
\end{equation}
this expression can be further simplified to: 
\begin{align}
G_{p}^{(N)} & =\sum_{j=0}^{\infty}\frac{\boldsymbol{n}^{p+j\mathcal{\mathcal{M}}}}{\left(p+j\mathcal{\mathcal{M}}\right)!}.\label{eq:phase-symmetry-inner-product}
\end{align}

A relationship with the hypergeometric function is obtained from Eq
(\ref{eq:phase-symmetry-inner-product}), by introducing $z\equiv(\boldsymbol{n}/\mathcal{\mathcal{\mathcal{M}}})^{\mathcal{\mathcal{\mathcal{M}}}}$.
After the variable change, 
\begin{equation}
G_{p}^{(N)}=\frac{\boldsymbol{n}_{p}}{p!}\sum_{j=0}^{\infty}\frac{p!z^{j}\mathcal{\mathcal{M}}^{j\mathcal{\mathcal{M}}}}{\left(p+j\mathcal{\mathcal{M}}\right)!}.
\end{equation}
The ratio of successive terms in the series can be used to obtain
the sum of terms as the hypergeometric series $\,_{1}F_{\mathcal{\mathcal{M}}}$:
\begin{align}
G_{p}^{(N)} & =(\boldsymbol{n}_{p}/p!)\,_{1}F_{\mathcal{\mathcal{M}}}(1;(p+1)/\mathcal{\mathcal{M}},\nonumber \\
 & ...(p+\mathcal{\mathcal{M}})/\mathcal{\mathcal{M}};(\boldsymbol{n}/\mathcal{\mathcal{M}})^{\mathcal{\mathcal{M}}}).
\end{align}
As above, analogous equations are obtained for the ``simple'' weight
when $\boldsymbol{n}=\bar{\boldsymbol{n}}=\left|\boldsymbol{\alpha}\right|^{2}.$

\subsection{Weight examples with a phase symmetry}

Using the inner products given above, three examples for a phase symmetry,
depending on $\mathcal{M}$ are as follows:
\begin{itemize}
\item $\mathcal{\mathcal{\mathcal{M}}}=1$: In this case there is single
eigenvalue, i.e. no superpositions, so $\bm{U}=\bm{I}$. This gives
the standard positive-P result, 
\begin{align}
G^{(N)} & =\exp(n)
\end{align}
\item $\mathcal{\mathcal{\mathcal{M}}}=2$:  This case applies to parity
symmetry, and corresponds to even and odd Schrödinger cat superpositions,
with:
\begin{align}
G_{p}^{(N)} & =\left[\cosh(n),\sinh(n)\right]_{p}
\end{align}
\item $\mathcal{\mathcal{\mathcal{M}}}\rightarrow\infty$: This gives the
limit of a continuous symmetry, in which case one obtains a number
state:
\begin{align}
G_{p}^{(N)} & =\frac{n^{p}}{p!}\label{eq:number-symmetry-normalisation}
\end{align}
\end{itemize}

\section{Dynamics and observables\label{sec:Dynamics-and-operator}}

A fundamental component of any phase-space representation are the
operator identities. These are required to map dynamical equations
of the density matrix onto Fokker-Planck equations (FPEs), and to
compute observables.

Previous techniques \citep{Carusotto:2001} for projected dynamical
equations with bosons used a phase symmetry with $\mathcal{\mathcal{M}}\rightarrow\infty$,
and only one projected eigenvalue, a Bloch state. Here, we investigate
the differential identities for the complete, positive Hilbert space
representation in more general cases. 

In this section, we summarize the matrix-P operator identities. Detailed
calculations of each identity can be found in Appendix C. 

\subsection{Dynamical equations }

To obtain dynamics in phase-space, one starts with a Liouville equation.
This could be for evolution with a Hamiltonian, where there is a commutator:
\begin{equation}
\frac{d\hat{\rho}}{dt}=-i\left[\hat{H},\hat{\rho}\right],\label{eq:Hamiltonian_evolution}
\end{equation}
 or for a master equation with open systems. For computing ground
states and canonical or grand canonical thermal density matrices \citep{Deuar:2009}
with a chemical potential $\mu$, one finds an anticommutator:
\begin{equation}
\frac{d\hat{\rho}}{d\beta}=-\left[\hat{H}-\mu\hat{N},\hat{\rho}\right]_{+}.\label{eq: Grand_canonical}
\end{equation}

All three cases have the structure:
\begin{equation}
\frac{d\hat{\rho}}{dt}=\mathcal{L}\left[\hat{\rho}\right]=\mathcal{L}_{\mu\nu}\hat{O}_{\mu}\hat{\rho}\hat{O}_{\nu},
\end{equation}
where we use an Einstein summation convention to sum over the Liouville
operator terms. Substituting the matrix representation Eq.(\ref{eq:Matrix_representation})
gives:
\begin{equation}
\int\dot{P}_{\mathcal{M}}(\vec{\lambda},t)\hat{\Lambda}_{\mathcal{M}}(\vec{\lambda})\text{d}\vec{\lambda}=\int P_{\mathcal{M}}(\vec{\lambda},t)\mathcal{L}_{\mu\nu}\hat{O}_{\mu}\hat{\Lambda}_{\mathcal{M}}(\vec{\lambda})\hat{O}_{\nu}\text{d}\vec{\lambda},
\end{equation}
where $\dot{P}_{\mathcal{M}}(\vec{\lambda},t)=\frac{d}{dt}P_{\mathcal{M}}(\vec{\lambda},t)$. 

Using the operator identities derived below, the operator terms of
form $\hat{O}_{i}\hat{\Lambda}_{\mathcal{M}}(\vec{\lambda})\hat{O}_{j}$
are mapped into differential operators, denoted $\mathcal{D}_{\mu\nu}$,
on phase-space, so that:
\begin{equation}
\hat{O}_{\mu}\hat{\Lambda}_{\mathcal{M}}(\vec{\lambda})\hat{O}_{\nu}=\mathcal{D}_{\mu\nu}\hat{\Lambda}_{\mathcal{M}}(\vec{\lambda}).\label{eq:general_differential operator mapping}
\end{equation}
Provided the phase-space boundary terms vanish, one can use partial
integration to transform Eq.(\ref{eq:general_differential operator mapping})
into an adjoint form $\tilde{\mathcal{D}}_{\mu\nu}$ acting on $P_{\mathcal{M}}(\vec{\lambda},t)$,
so that there is a solution which satisfies:
\begin{equation}
\dot{P}_{\mathcal{M}}(\vec{\lambda},t)=\mathcal{L}_{\mu\nu}\tilde{\mathcal{D}}_{\mu\nu}P_{\mathcal{M}}(\vec{\lambda},t).
\end{equation}

If the resulting generalized FPE has no higher than second order derivatives,
and has a positive definite diffusion, it then has a stochastic equivalent
which can be efficiently randomly sampled \citep{Gardiner_Book_SDE}.
A complex distribution is also possible, and is useful in finding
exact solutions \citep{drummond1980quantum,drummond1981nonequilibrium}.

\subsection{State ket identities}

We start with the simplest identities; operators acting on the projected
coherent states, Eq.(\ref{eq:projected-state}). These form the basis
for the kernel identities, which we treat below. 

The projected operator identities stem from the fundamental coherent
state identities \citep{Bargmann:1961,Glauber_1963_P-Rep}. Upon defining
coherent derivatives as  $\partial_{j}=\partial/\partial\alpha_{j}$,
these are 
\begin{align}
\hat{a}_{j}\left\Vert \boldsymbol{\alpha}\right\rangle  & =\alpha_{j}\left\Vert \boldsymbol{\alpha}\right\rangle \nonumber \\
\hat{a}_{j}^{\dagger}\left\Vert \boldsymbol{\alpha}\right\rangle  & =\partial_{j}\left\Vert \boldsymbol{\alpha}\right\rangle .\label{eq:coherent_operator_identities}
\end{align}

For projected coherent states, the complete symmetry identities are:
\begin{align}
\hat{a}_{k}\underline{\left\Vert \boldsymbol{\alpha}\right\rangle } & =\alpha_{k}\underline{\sigma}_{k}\underline{\left\Vert \boldsymbol{\alpha}\right\rangle }\nonumber \\
\hat{a}_{k}^{\dagger}\underline{\left\Vert \boldsymbol{\alpha}\right\rangle } & =\partial_{k}\underline{\sigma}_{k}^{T}\underline{\left\Vert \boldsymbol{\alpha}\right\rangle },\label{eq:ket-identities-2}
\end{align}
where $\underline{\left\Vert \boldsymbol{\alpha}\right\rangle }=\left[\left\Vert \boldsymbol{\alpha}\right\rangle _{0},\ldots,\left\Vert \boldsymbol{\alpha}\right\rangle _{\mathcal{M}-1}\right]^{T}$.
The matrices $\underline{\sigma}_{k}$ reduce the global eigenvalues
$\tilde{p}$ by the corresponding quantum number vector $\tilde{h}_{k}$
for a particle in mode $k$, while $\underline{\sigma}_{k}^{T}$ does
the opposite, increasing the eigenvalues (see Appendix C). 

As in the Appendix, the identities assume that the modes are chosen
such that the symmetry generators $\tilde{\bm{g}}$ are diagonal in
the mode index, and are written as a vector $\tilde{h}_{k}$ over
the group transformation index $\nu$, for mode $k$.

\subsubsection{Ket phase symmetry example}

As an example, we treat a phase symmetry for the three $\mathcal{M}$
case used above. Details can be found in the Appendix. 
\begin{itemize}
\item $\mathcal{M}=1$: There is no symmetry projection, so $\underline{\sigma}_{k}=1$.
The identities are the coherent state identities of Eq.(\ref{eq:coherent_operator_identities}).
\item $\mathcal{M}=2$: The $\underline{\sigma}_{k}$ matrix is a Pauli
matrix, $\underline{\sigma}_{k}=\underline{\sigma}_{k}^{T}=\underline{\sigma}^{x}$.
In this case the identities Eq.(\ref{eq:ket-identities-2}) change
by $\pm1$ for any mode. However, due to parity conservation, the
quadratic identities reduce to the standard coherent state case, since
$\left(\underline{\sigma}^{x}\right)^{2}=\underline{I}$:
\begin{align}
\hat{a}_{j}^{2}\underline{\left\Vert \boldsymbol{\alpha}\right\rangle } & =\alpha_{j}^{2}\underline{\left\Vert \boldsymbol{\alpha}\right\rangle }\nonumber \\
\hat{a}_{i}^{\dagger}\hat{a}_{j}\underline{\left\Vert \boldsymbol{\alpha}\right\rangle } & =\alpha_{j}\partial_{i}\underline{\left\Vert \boldsymbol{\alpha}\right\rangle }\nonumber \\
\hat{a}_{j}^{\dagger2}\underline{\left\Vert \boldsymbol{\alpha}\right\rangle } & =\partial_{j}^{2}\underline{\left\Vert \boldsymbol{\alpha}\right\rangle }.
\end{align}
\item $\mathcal{M}=\infty$: One still has $\underline{\sigma}\underline{\sigma}^{T}=\underline{I}$,
so just as in the standard positive-P case
\begin{equation}
\hat{a}_{i}^{\dagger}\hat{a}_{j}\underline{\left\Vert \boldsymbol{\alpha}\right\rangle }=\alpha_{j}\partial_{i}\underline{\left\Vert \boldsymbol{\alpha}\right\rangle }.\label{eq:quad_ket_ident_phase}
\end{equation}
Note that this is only true for a phase-symmetry whose eigenvalues
are independent of the mode index. 
\end{itemize}

\subsection{Kernel identities}

The state ket identities are used to derive the kernel identities,
which in turn are needed to map Liouville and master equations to
FPEs. For simplicity, we treat the kernel matrix element identities
here, while the full identities are obtained in the Appendix. 

For normal ordering, the identity for a general global symmetry
is:
\begin{align}
\hat{a}_{k}\hat{\Lambda}_{\tilde{p}\tilde{q}} & =\alpha_{k}\Sigma_{k}^{\tilde{p}\tilde{p}'}\left(\vec{\alpha}\right)\hat{\Lambda}_{\tilde{p}'\tilde{q}}\nonumber \\
\hat{\Lambda}_{\tilde{p}\tilde{q}}\hat{a}_{k}^{\dagger} & =\beta_{k}\Sigma_{k}^{\tilde{q}\tilde{q}'}\left(\vec{\alpha}'\right)\hat{\Lambda}_{\tilde{p}\tilde{q}'}.\label{eq:normal-ordered identities}
\end{align}
For anti-normal ordering, we define $\partial_{j}^{+}=\partial/\partial\beta_{j}$,
so that
\begin{align}
\hat{a}_{k}^{\dagger}\hat{\Lambda}_{\tilde{p}\tilde{q}} & =\bar{\Sigma}_{k}^{\tilde{p}\tilde{p}'}\left(\vec{\alpha}\right)\left(\partial_{k}+\frac{\partial w_{\tilde{p}'\tilde{q}}}{\partial\alpha_{k}}\right)\hat{\Lambda}_{\tilde{p}'\tilde{q}}\nonumber \\
\hat{\Lambda}_{\tilde{p}\tilde{q}}\hat{a}_{k} & =\bar{\Sigma}_{k}^{\tilde{q}\tilde{q}'}\left(\vec{\alpha}'\right)\left(\partial_{k}^{+}+\frac{\partial w_{\tilde{p}\tilde{q}'}}{\partial\alpha_{k}}\right)\hat{\Lambda}_{\tilde{p}\tilde{q}'}.\label{eq:anti-normal identities}
\end{align}

Here, we introduce coefficients
\begin{align}
\Sigma_{k}^{\tilde{p}\tilde{p}'}\left(\vec{\alpha}\right) & \equiv\sigma_{k}^{\tilde{p}\tilde{p}'}e^{w_{\tilde{p}'}\left(\vec{\alpha}\right)-w_{\tilde{p}}\left(\vec{\alpha}\right)}\nonumber \\
\bar{\Sigma}_{k}^{\tilde{p}\tilde{p}'}\left(\vec{\alpha}\right) & \equiv\bar{\sigma}_{k}^{\tilde{p}\tilde{p}'}e^{w_{\tilde{p}'}\left(\vec{\alpha}\right)-w_{\tilde{p}}\left(\vec{\alpha}\right)},
\end{align}
that are transform matrices defined using $\sigma_{k}^{\tilde{p}\tilde{p}'}$
and $\bar{\sigma}_{k}^{\tilde{p}\tilde{p}'}$, the Fourier transforms
of the unitary matrix (see Appendix C for definitions). Just like
the corresponding quantum operators, $\Sigma$ annihilates quantum
numbers while $\bar{\Sigma}$ creates them. There is a difference,
which is that our definition makes the quantum numbers a cyclic group,
which can be understood more clearly in a phase symmetry example.

\subsubsection{Kernel phase symmetry example}

In the phase symmetry case, $h_{k}=1$ and the transform matrix is
cyclic, such that following Eq.(\ref{eq:weight_factorization}) and
Eq.(\ref{eq:periodic_kronecker}), $\sigma_{k}^{pp'}=\delta_{p-1,p'}^{(\mathcal{M})}$
and $\Sigma_{k}^{pp'}\left(\vec{\alpha}\right)=\sqrt{T_{p}}\sigma_{k}^{pp'}$,
where 
\begin{equation}
T_{p}=\frac{G_{p-1}\left(\vec{\alpha}\right)}{G_{p}\left(\vec{\alpha}\right)},
\end{equation}
while $T'_{q}=G_{q-1}\left(\vec{\alpha}'\right)/G_{q}\left(\vec{\alpha}'\right)$.
Since the eigenvalue index of the states change with application of
the creation and annihilation operators, the kernel operator must
be renormalized. This is achieved through $T_{p}$, which are elements
of a $\mathcal{M}\times\mathcal{M}$ diagonal $(T_{pq}=\delta_{pq}T_{p})$
renormalization matrix $\underline{T}$. We recall that the indices
are cyclic, so $G_{-1}=G_{\mathcal{M}-1}$. 

The normally ordered identities therefore simplify to 
\begin{align}
\hat{a}_{j}\hat{\Lambda}_{pq} & =\alpha_{j}\sqrt{T_{p}}\Lambda_{p-1,q}\nonumber \\
\hat{\Lambda}_{pq}\hat{a}_{j}^{\dagger} & =\beta_{j}\sqrt{T'_{q}}\hat{\Lambda}_{p,q-1}.\label{eq:Phase-ident-n-ordered}
\end{align}

From Eq.(\ref{eq:anti-normal identities}), the partial derivative
means anti-normal ordered identities require one to specify a weight
function. For the normal weight case, $T_{p}=T'_{p}$ and one has
\begin{align}
a_{j}^{\dagger}\hat{\Lambda}_{pq} & =\frac{1}{\sqrt{T_{p+1}}}\left(\partial_{j}+\frac{\beta_{j}}{2}\left[T_{p+1}+T_{q}\right]\right)\Lambda_{p+1,q}\nonumber \\
\hat{\Lambda}_{pq}a_{j} & =\frac{1}{\sqrt{T_{q+1}}}\left(\partial_{j}^{+}+\frac{\alpha_{j}}{2}\left[T_{p}+T_{q+1}\right]\right)\Lambda_{p,q+1}.
\end{align}
 Meanwhile, for the simple weight with $T_{p}\neq T'_{p}$
\begin{align}
a_{j}^{\dagger}\hat{\Lambda}_{pq} & =\frac{1}{\sqrt{T_{p+1}}}\left(\partial_{j}+\frac{\alpha_{j}^{*}}{2}T_{p+1}\right)\Lambda_{p+1,q}\nonumber \\
\hat{\Lambda}_{pq}a_{j} & =\frac{1}{\sqrt{T'_{q+1}}}\left(\partial_{j}^{*}+\frac{\alpha_{j}}{2}T'_{q+1}\right)\Lambda_{p,q+1}.
\end{align}
In both cases, one has that $\bar{\sigma}_{k}^{pp'}=\delta_{p+1,p'}^{(\mathcal{M})}$. 

\subsubsection{Specific cases}

To give specific examples of this:
\begin{itemize}
\item $\mathcal{M}=2$: For the normal weight, $T_{0}=t\equiv\tanh\left(n\right)$
and $T_{1}=1/t$ giving 
\begin{equation}
\underline{T}=\underline{T}'=\left[\begin{array}{cc}
t & 0\\
0 & 1/t
\end{array}\right],\label{eq:M_2_phase_sym_renorm_matrix}
\end{equation}
while $T_{0}=\bar{t}\equiv\tanh\left(\left|\alpha\right|^{2}\right)$
and $T'_{0}=\bar{t}'\equiv\tanh\left(\left|\beta\right|^{2}\right)$
for the simple weight, such that 
\begin{align}
\underline{T} & =\left[\begin{array}{cc}
\bar{t} & 0\\
0 & 1/\bar{t}
\end{array}\right]\\
\underline{T}' & =\left[\begin{array}{cc}
\bar{t}' & 0\\
0 & 1/\bar{t}'
\end{array}\right].
\end{align}
\item $\mathcal{M}\rightarrow\infty$: From Eq.(\ref{eq:number-symmetry-normalisation})
$T_{p}=T'_{p}=p/n$ for the normal weight, while $T_{p}=p/\left|\alpha\right|^{2}$,
$T'_{p}=p/\left|\beta\right|^{2}$ for the simple weight.
\end{itemize}

\subsubsection{Complex gauge identities}

There is an additional matrix identity 
\begin{equation}
\left(\Omega_{\tilde{p}\tilde{q}}\frac{\partial}{\partial\Omega_{\tilde{p}\tilde{q}}}-1\right)\hat{\Lambda}_{\mathcal{M}}=0.
\end{equation}

As in the gauge-P representation, this identity modifies the resulting
FPEs to include gauges in either the drift or diffusion terms. The
type of gauge one wishes to use depends on whether the resulting stochastic
equation has singularities, in which case drift gauges are used \citep{Deuar:2002},
or whether large sampling errors are present, where diffusion gauges
are useful. 

Treating systems with variable $\Omega$ will be addressed in future
work, and we find that these terms typically appear in cases of non-hermitian
evolution, where the density matrix trace changes, either in time
or in imaginary time. 

\subsection{Observables}

The above identities can be used to calculate expectation values of
observables $\hat{O}$, which have corresponding c-number matrix functions
$\underline{O}\left(\vec{\lambda}\right)$, such that, provided $\text{Tr}\left[\hat{\rho}\right]=1$,
\begin{equation}
\left\langle \hat{O}\right\rangle =\int\text{Tr}_{\mathcal{M}}\left[\underline{\Omega}\underline{O}\left(\vec{\lambda}\right)\right]P_{\mathcal{M}}(\vec{\lambda},t)\text{d}\vec{\lambda}.
\end{equation}

We first obtain the general expression for a normally ordered moment.
This gives the correlations in Glauber's coherence theory \citep{Glauber1963_CoherentStates,Glauber1963photon,Glauber_1963_P-Rep}.
It is a moment with $m$ annihilation and $\bar{m}$ creation operators.
Using the cyclic properties of the operator trace, and expanding the
quantum density matrix in a matrix P-representation from Eq.(\ref{eq:Matrix_representation}):
\begin{align}
\left\langle \hat{O}\right\rangle  & =\text{Tr}\left[\hat{\rho}\hat{a}_{k_{1}}^{\dagger}\ldots\hat{a}_{k_{\bar{m}}}^{\dagger}\hat{a}_{j_{m}}\ldots\hat{a}_{j_{1}}\right]\nonumber \\
 & =\text{Tr}\left[\int P_{\mathcal{M}}(\vec{\lambda},t)\text{Tr}_{\mathcal{M}}\left[\underline{\Omega}\hat{a}_{j_{m}}\ldots\hat{\underline{\Lambda}}(\vec{\alpha})\hat{a}_{k_{1}}^{\dagger}\ldots\right]\text{d}\vec{\lambda}\right].
\end{align}
From the normally-ordered identities of Eq.(\ref{eq:normal-ordered identities}),
recalling that traces and integrals are linear and can be interchanged
if uniformly convergent and the identity Eq.(\ref{eq:kernel trace result}),
we obtain:
\begin{align}
\left\langle \hat{O}\right\rangle  & =\int\alpha_{j_{1}}\ldots\beta_{k_{1}}\ldots\text{Tr}_{\mathcal{M}}\left(\underline{\Sigma^{\prime T}}{}_{k_{\bar{m}}}\ldots\underline{\Omega}\right.\nonumber \\
 & \left.\underline{\Sigma}{}_{j_{1}}\ldots\right)P_{\mathcal{M}}(\vec{\lambda},t)\text{d}\vec{\lambda}.\label{eq:Matrix-P-correlations}
\end{align}

Since $\underline{\Omega}$ is present, expectation values depend
on the additional weights \citep{Deuar:2002}. 

\subsubsection{Observable phase symmetry example}

To give a more detailed example of Eq.(\ref{eq:Matrix-P-correlations}),
we treat phase symmetry moments with $\mathcal{N}=1$ and $\mathcal{M}=2$.
Using the identities above gives the result for parity symmetry with
normal weights,
\begin{align}
\left\langle :\hat{n}_{j}^{m}:\right\rangle  & =\int\left(\alpha_{j}\beta_{j}\right)^{m}\text{Tr}_{\mathcal{M}}\left[\underline{T}^{\pi_{m}}\underline{\Omega}\right]P_{\mathcal{M}}(\vec{\lambda},t)\text{d}\vec{\lambda},\label{eq:moment-2}
\end{align}
where $\pi_{m}=\left[1-(-1)^{m}\right]/2$, so $\pi_{m}=0$ (1) for
even (odd) $m$.

\subsubsection*{Proof:}

Letting $\hat{O}=:\hat{n}_{j}^{m}:,$ from the definition of the matrix-P
representation and the standard form of a quantum expectation value,
one has that:
\begin{align}
\left\langle :\hat{n}_{j}^{m}:\right\rangle  & =\int P_{\mathcal{M}}(\vec{\lambda},t)\text{Tr}_{\mathcal{M}}\left[\underline{\Omega}\text{Tr}\left(\hat{\underline{\Lambda}}(\vec{\alpha})\hat{O}\right)\right]\text{d}\vec{\lambda}.
\end{align}

From the cyclic properties of the operator trace:
\begin{align}
\text{Tr}\left(\hat{\underline{\Lambda}}(\vec{\alpha})\hat{O}\right) & =\text{Tr}\left(a_{j}^{m}\hat{\underline{\Lambda}}(\vec{\alpha})a_{j}^{\dagger m}\right).
\end{align}
Using the identities Eq.(\ref{eq:Phase-ident-n-ordered}) and Eq.(\ref{eq:kernel trace result}),
the operator trace becomes
\begin{align}
\text{Tr}\left(a_{j}^{m}\hat{\underline{\Lambda}}(\vec{\alpha})a_{j}^{\dagger m}\right) & =\alpha_{j}^{m}\beta_{j}^{m}\left(\underline{T}^{1/2}\underline{\sigma}\right){}^{m}\left(\underline{\sigma}\underline{T}^{1/2}\right){}^{m}.
\end{align}

The renormalization matrices are given in Eq.(\ref{eq:M_2_phase_sym_renorm_matrix}).
If $m$ is even, one has $\left(\underline{T}^{1/2}\underline{\sigma}\right){}^{m}=\left(\underline{\sigma}\underline{T}^{1/2}\right){}^{m}=\underline{I}$.
If $m$ is odd, we can always rewrite the powers in terms of an even
$m$ value and the central $m=1$ result, which gives 
\begin{equation}
\underline{T}^{1/2}\left(\underline{\sigma}\right)^{2}\underline{T}^{1/2}=\underline{T}.
\end{equation}
Therefore, these two results immediately prove Eq.(\ref{eq:moment-2}). 

The applications presented in the subsequent section require multimode
moments. These are readily obtained by generalizing the above to products
of moments with more than one mode. If $\bm{m}=\left[m_{1},m_{2},\ldots\right]$,
$m=\sum_{i}m_{i}$ and $n_{i}=\alpha_{i}\beta_{i}$, then one finds
that:
\begin{equation}
\left\langle :\hat{n}_{1}^{m_{1}}\hat{n}_{2}^{m_{2}}\ldots:\right\rangle =\int\left(n_{1}^{m_{1}}n_{2}^{m_{2}}\ldots\right)\text{Tr}_{\mathcal{M}}\left[\underline{T}^{\pi_{m}}\underline{\Omega}\right]P_{\mathcal{M}}(\vec{\lambda})\text{d}\vec{\lambda}.
\end{equation}
Grouping photon numbers into subsets $S_{j}$ so that
\begin{align}
\hat{N}_{j} & =\sum_{i\in S_{j}}\hat{n}_{i},
\end{align}
with $N_{j}=\sum_{i\in S_{j}}n_{i}$ being the corresponding phase-space
variable, leads to a general normally-ordered multimode grouped moment:
\begin{align}
\left\langle \hat{O}\right\rangle  & =\left\langle :\hat{N}_{1}^{m_{1}}\hat{N}_{2}^{m_{2}}\ldots:\right\rangle \nonumber \\
 & =\int\left(N_{1}^{m_{1}}N_{2}^{m_{2}}\ldots\right)\text{Tr}_{\mathcal{M}}\left[\underline{T}^{\pi_{m}}\underline{\Omega}\right]P_{\mathcal{M}}(\vec{\lambda})\text{d}\vec{\lambda}.\label{eq:multimode grouped moment}
\end{align}

As an example of an even moment, for which the extra factor of $\underline{T}$
does not enter, one has that:
\begin{align}
\left\langle :\hat{N}_{i}^{2}:\right\rangle  & =\int N_{i}^{2}\text{Tr}_{\mathcal{M}}\left[\underline{\Omega}\right]P_{\mathcal{M}}(\vec{\lambda})\text{d}\vec{\lambda}.\label{eq:moment-1}
\end{align}

\section{Gaussian boson sampling\label{sec:Applications}}

In this section, we show how to apply the matrix-P representation
to the validation of GBS quantum computers, as initially presented
in \citep{drummond2026matrix}. The task of GBS devices is to generate
a desired random distribution by transmitting photons in a squeezed
vacuum state into a linear optical network of beamsplitters and phase-shifters.
These linearly transform the inputs into outputs which are measured
by photo-detectors, to give each random number sample. 

The probability of an output sample is either a matrix Hafnian \citep{Hamilton2017gaussian},
for photon-number resolving (PNR) detectors or the Torontonian \citep{quesada2018gaussian},
for threshold detectors. Although both functions are \#P-hard to compute,
we focus on PNR implementations. Since large-scale probabilities cannot
be either measured or computed, our goal is to efficiently compute
the probabilities of observable marginals and binned counts.

The purpose of validation studies is to obtain quantitative information
about QC errors. This allows an analysis of whether quantum computational
advantage has been achieved. To do this, one compares QC errors from
experiments \citep{zhongPhaseProgrammableGaussianBoson2021,madsenQuantumComputationalAdvantage2022,deng2023gaussian,liu2025robust}
with those of classical approximate count generators \citep{villalonga2021efficient,oh2024classical,dodd2025fast,goodman2026gaussian}.
These algorithms approximately replicate the GBS task, as exact methods
\citep{bulmerBoundaryQuantumAdvantage2022a} are not scalable. While
this is outside our present scope, we note that phase-space methods
can provide a route to these approximate algorithms \citep{MartinezCifuentes2023classicalmodelsmay,goodman2026gaussian}.

\subsection{Photon-counting probabilities}

If a set $S$ of photo-detectors has disjoint subsets $S_{j},$ the
projection operator for the photon number vector $\bm{c}=[c_{1},\dots,c_{M}]$
of measurements in $S_{j}$ is
\begin{equation}
\hat{\mathcal{G}}\left(\boldsymbol{c}\right)=\bigotimes_{i\in S_{j}}\frac{1}{c_{i}!}:(\hat{n}_{i})^{c_{i}}e^{-\hat{n}_{i}}:.\label{eq:Number-projector}
\end{equation}
Defining $m=\sum_{i\in S}c_{i}$, the quantum projection operator
for observing $m$ \textbf{total }counts in the set $S=\{1,\dots,M\}$
is obtained by summing over all combinations of output count patterns
such that: 
\begin{equation}
\hat{\mathcal{G}}_{S}\left(\boldsymbol{m}\right)=\sum_{m=\sum c_{i}}\prod_{i\in S}\left(\frac{1}{c_{i}!}:\hat{n}_{i}^{c_{i}}e^{-\hat{n}_{i}}:\right).
\end{equation}

From the multinomial theorem this can be rewritten as:
\begin{equation}
\hat{\mathcal{G}}_{S}\left(m\right)=\frac{1}{m!}:\hat{N}^{m}e^{-\hat{N}}:,\label{eq:Number-projector-1}
\end{equation}
where the grouped number operator is defined as $\hat{N}=\sum_{i\in S}\hat{n}_{i}.$

We call expectation values of Eq.(\ref{eq:Number-projector-1}) the
total count distribution. For uniform pure squeezed state photons
this distribution can be calculated exactly in both lossless \citep{huangPhotoncountingStatisticsMultimode1989,zhuPhotocountDistributionsContinuouswave1990,Hamilton2017gaussian}
and uniform loss \citep{deshpandeQuantumComputationalAdvantage2022a}
photonic networks. 

In this work we focus on comparisons with the total count distribution
as an example. However a similar formula therefore applies if the
pattern $\bm{m}=[m_{1},m_{2},\dots]$ describes the total numbers
recorded from a set of bins, where $\bm{S}=\left[S_{1},S_{2}\ldots\right]$,
and now
\begin{equation}
\hat{\mathcal{G}}_{\bm{S}}\left(\boldsymbol{m}\right)=\bigotimes_{j}\frac{1}{m_{j}!}:\hat{N}_{j}{}^{m_{j}}e^{-\hat{N}_{j}}:.
\end{equation}

The exponential prefactor can be separated into even and odd powers.
Therefore, defining the total sum over bins as $\hat{N}_{S}=\sum_{j}\hat{N}_{j}$,
the binned number projector can now be rewritten as:
\begin{equation}
\hat{\mathcal{G}}_{\bm{S}}\left(\boldsymbol{m}\right)=:\left[\mathcal{C}\left(\hat{N}_{\bm{S}}\right)-\mathcal{S}\left(\hat{N}_{\bm{S}}\right)\right]\bigotimes_{j}\frac{\hat{N}_{j}^{m_{j}}}{m_{j}!}:,\label{eq:binned number projector}
\end{equation}
where $\mathcal{C}=\cosh$ (even) and $\mathcal{S}=\sinh$ (odd).
The set $\bm{S}$ does not have to include all possible modes, and
we can take advantage of this to include a set of reservoir modes
that model undetected photons, or losses.

\subsection{Photon counting with parity symmetry}

Pure squeezed vacuum state photons are always generated in correlated
pairs. Therefore, we can simulate the photon counting distributions
using an $\mathcal{M}=2$ parity symmetric matrix-P representation
with normal weights. This gives the result for the parity representation
with normal weights, that, if $m=\sum_{j}m_{j},$ $N_{S}=\sum_{j}N_{j}$,
and the parity is $\pi_{m}=\left[1-(-1)^{m}\right]/2$, $\bar{\pi}_{m}\equiv1-\pi_{m}$,
then from Eq (\ref{eq:multimode grouped moment}) the c-number matrix
function corresponding to $\hat{\mathcal{G}}_{\bm{S}}\left(\boldsymbol{m}\right)$
is: 
\begin{equation}
\mathcal{G}_{\bm{S}}\left(\bm{m}\right)=\left[\mathcal{C}\left(N_{S}\right)\underline{T}^{\pi_{m}}-\mathcal{S}\left(N_{S}\right)\underline{T}^{\bar{\pi}_{m}}\right]\prod_{j\in S}\frac{N_{i}^{m_{j}}}{m_{j}!}.\label{eq:photon-counting-identity-1}
\end{equation}

Since $\bm{T}$ is a diagonal matrix, and $\pi_{m}=[0,1]$, we note
that:
\begin{align}
\underline{T}^{\pi_{m}}+\underline{T}^{\bar{\pi}_{m}} & =\underline{1}+\underline{T},\\
\underline{T}^{\pi_{m}}-\underline{T}^{\bar{\pi}_{m}} & =\left(\underline{T}-\underline{1}\right)\left(\pi_{m}-\bar{\pi}_{m}\right).\nonumber 
\end{align}
Hence, this can be rewritten as: 
\begin{equation}
\mathcal{G}_{\bm{S}}\left(\bm{m}\right)=\left[\begin{array}{cc}
\mathcal{G}_{S}^{0}\left(N_{s},m\right) & 0\\
0 & \mathcal{G}_{S}^{1}\left(N_{s},m\right)
\end{array}\right]\prod_{j\in S}\frac{N_{i}^{m_{j}}}{m_{j}!}e^{-N_{i}},
\end{equation}

where:
\begin{align}
\mathcal{G}_{S}^{0}\left(N_{s},m\right) & =\frac{1}{2}\left(e^{2N_{s}}\left(t-1\right)\left(\pi_{m}-\bar{\pi}_{m}\right)+\left(t+1\right)\right)\\
\mathcal{G}_{S}^{1}\left(N_{s},m\right) & =\frac{1}{2}\left(e^{2N_{s}}\left(1/t-1\right)\left(\pi_{m}-\bar{\pi}_{m}\right)+\left(1/t+1\right)\right)\nonumber 
\end{align}

This can be simplified further to give the prefactor $\mathcal{G}^{0}$
for an even parity projection like a squeezed state, which is:
\begin{equation}
\mathcal{G}_{S}^{0}\left(N_{s},m\right)=\left(\frac{1-\left(\pi_{m}-\bar{\pi}_{m}\right)e^{2\left(N_{s}-n\right)}}{1+e^{-2n}}\right),
\end{equation}
where, as elsewhere, $n=\sum\alpha_{i}\beta_{i}$. 

Although it might seem initially that one is restricted to lossless
cases to obtain parity conservation, this is not true. Loss channels
can be modeled as additional reservoir modes. These are not measured,
so the set of measured channels $S$ does not include them. The coupling
of the non-reservoir inputs to the reservoirs leads to a non-hermitian
coupling for the monitored channels. While $n$ gives the total photon
number for all modes including reservoirs, $N_{s}$ includes only
the measured channels.

\subsection{Linear network with phase symmetry}

Applications of phase-space methods typically require dynamical equations.
These must not have exponential complexity, otherwise they would be
computationally intractable at large mode number. To demonstrate this,
we treat dynamics obtained with a linear optical network, where the
Hamiltonian is $\hat{H}=\omega_{ij}\hat{a}_{i}^{\dagger}\hat{a}_{j}$
$(\hbar=1)$. This has a global phase symmetry for any value of $\mathcal{M}$,
leading to conservation of parity and particle number. 

The evolution equation is: 
\begin{align}
\dot{\rho} & =-i\omega_{ij}\left[\hat{a}_{i}^{\dagger}\hat{a}_{j},\hat{\rho}\right].
\end{align}
Because the Hamiltonian has a phase symmetry, it doesn't change projected
eigenvalues. Applying the identities in the case of normal weights,
and noting the terms involving constants all cancel, the density matrix
equations are: 
\begin{align}
\dot{\rho} & =-i\omega_{ij}\int P_{\mathcal{M}}(\vec{\lambda},t)\text{Tr}_{\mathcal{M}}\left[\underline{\Omega}\left(\alpha_{j}\partial_{i}-\beta_{i}\partial_{j}^{+}\right)\hat{\underline{\Lambda}}(\vec{\alpha})\right]
\end{align}

After partial integration, we obtain an identical result for each
symmetry eigenvalue, giving:
\begin{equation}
\frac{\partial P_{\mathcal{M}}(\vec{\lambda},t)}{\partial t}=i\omega_{ij}\left[\frac{\partial}{\partial\alpha_{i}}\alpha_{j}-\frac{\partial}{\partial\beta_{i}}\beta_{j}\right]P_{\mathcal{M}}(\vec{\lambda},t).
\end{equation}
This matrix FPE is then solved by means of characteristics, with a
result that is similar to the corresponding operator solutions:
\begin{align}
\bm{\alpha}\left(t\right) & =e^{-i\underline{\omega}t}\bm{\alpha}\left(0\right)\nonumber \\
\bm{\beta}\left(t\right) & =e^{i\underline{\omega}t}\bm{\beta}\left(0\right).
\end{align}

In summary, for a linear network, the trajectories are transformed
by a unitary matrix, just as in the positive-P solutions \citep{dellios2021}.
This is because each component evolves independently of the other
terms. There are extra factors for nonlinear terms, but these are
not required here. Damping can be treated as well, by including a
coupling to a large mode reservoir initially in the vacuum state.
This is exactly equivalent to replacing $\underline{\omega}$ by a
non-hermitian transmission matrix that includes losses.

\subsection{Multimode squeezed state with parity symmetry}

To obtain the parity symmetry representation of a multimode squeezed
vacuum state $\left|\bm{r}\right\rangle $, a squeezed state can be
expanded as an integral over real coherent states \citep{adam1994complete},
in terms of an even function of $\bm{\alpha}$ defined as: 
\begin{equation}
\xi\left(\bm{\alpha}\right)=\prod_{j}C_{j}e^{-\alpha_{j}^{2}\coth\left(r_{j}\right)/2},
\end{equation}
where $C_{j}=\left[2\pi\sinh\left(r_{j}\right)\right]^{-1/2}$, for
real $r_{j}>0$. The state $\left|\bm{r}\right\rangle $ is an integral
over a real line in phase-space:
\begin{equation}
\left|\bm{r}\right\rangle =\int\xi\left(\bm{\alpha}\right)\left\Vert \boldsymbol{\alpha}\right\rangle d^{M}\bm{\alpha}.
\end{equation}
The positive-P distribution for pure squeezed states is then defined
on a $2M$ dimensional subspace \citep{drummondSimulatingComplexNetworks2022}
\begin{equation}
P(\vec{\alpha})=\prod_{j}c_{j}^{2}e^{-\left(\alpha_{j}^{2}+\beta_{j}^{2}\right)\coth\left(r_{j}\right)/2+n_{j}}\delta\left(\text{Im}\left(\vec{\alpha}\right)\right).
\end{equation}

Hence, for a multi-mode density matrix $\hat{\rho}=\left|\boldsymbol{r}\right\rangle \left\langle \boldsymbol{r}\right|$,
we have:
\begin{align}
\hat{\rho} & =\iint\xi\left(\bm{\alpha}\right)\xi\left(\bm{\beta}\right)\left\Vert \boldsymbol{\alpha}\right\rangle _{p}\left\langle \bm{\beta}\right\Vert _{p}d^{M}\bm{\alpha}d^{M}\bm{\beta}.
\end{align}
Using the even symmetry of $\xi\left(\bm{\alpha}\right)$, and noting
that for an $\mathcal{M}=2$ phase symmetry the projected state is
\begin{equation}
\left\Vert \bm{\alpha}\right\rangle _{p}=\frac{1}{2}\left(\left\Vert \bm{\alpha}\right\rangle +\left(-1\right)^{p}\left\Vert -\bm{\alpha}\right\rangle \right),
\end{equation}
 it follows that 
\begin{align}
\hat{\rho} & =\iint\xi\left(\bm{\alpha}\right)\xi\left(\bm{\beta}\right)\left\Vert \boldsymbol{\alpha}\right\rangle _{0}\left\langle \bm{\beta}\right\Vert _{0}d^{M}\bm{\alpha}d^{M}\bm{\beta}\nonumber \\
 & =\iint P\left(\vec{\lambda},t\right)\Lambda_{00}\left(\vec{\lambda}\right)d\vec{\lambda},
\end{align}
with the final result that:
\begin{equation}
P\left(\vec{\lambda}\right)=\xi\left(\bm{\alpha}\right)\xi\left(\bm{\beta}\right)e^{w_{00}\left(\vec{\alpha}\right)}\delta\left(\underline{\Omega}-\tilde{\underline{\Omega}}\right).
\end{equation}

For the parity matrix-P representation, only even terms are needed,
with a probability proportional to $n^{2p}$, giving a Gamma distribution
in the single-mode case. 

\subsection{Comparisons with exact solutions}

Validating quantum computing experiments is an important topic, since
useful computers must be both fast and precise. To enable practical
applications, and to determine whether there is a QC advantage it
is essential to validate the results, in order to know how accurate
the QC outputs are. 

A practical application of matrix P-representations is therefore to
efficiently compute observable moments of GBS experiments. As stated
above, for a lossless network with uniform pure squeezed state inputs
an exact total count distribution is known \citep{huangPhotoncountingStatisticsMultimode1989,zhuPhotocountDistributionsContinuouswave1990,Hamilton2017gaussian}.
This distribution contains distinct oscillations between even and
odd counts $m$, where the odd count probability is zero due to the
photons being generated in correlated pairs. This behavior is certainly
not just restricted to the special case that is analytically soluble,
because it is caused by the physics of squeezed vacuum states. These
states only have even numbers of photons, and provided there are no
losses, this property is independent of the number of squeezed inputs,
their amplitude, and any subsequent unitary transformations of the
modes.

\begin{figure}
\begin{centering}
\includegraphics[width=0.4\textwidth]{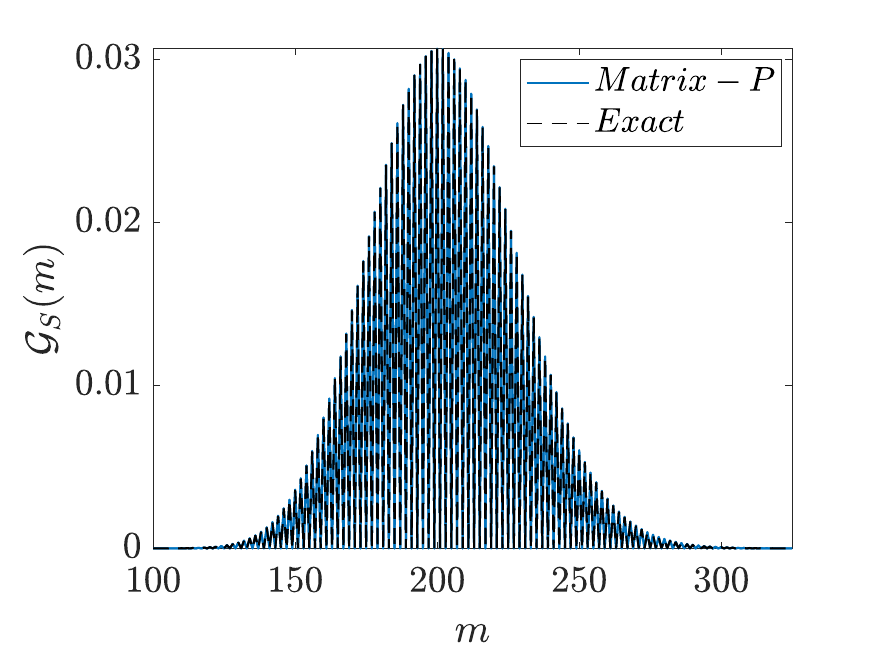}
\par\end{centering}
\begin{centering}
\includegraphics[width=0.4\textwidth]{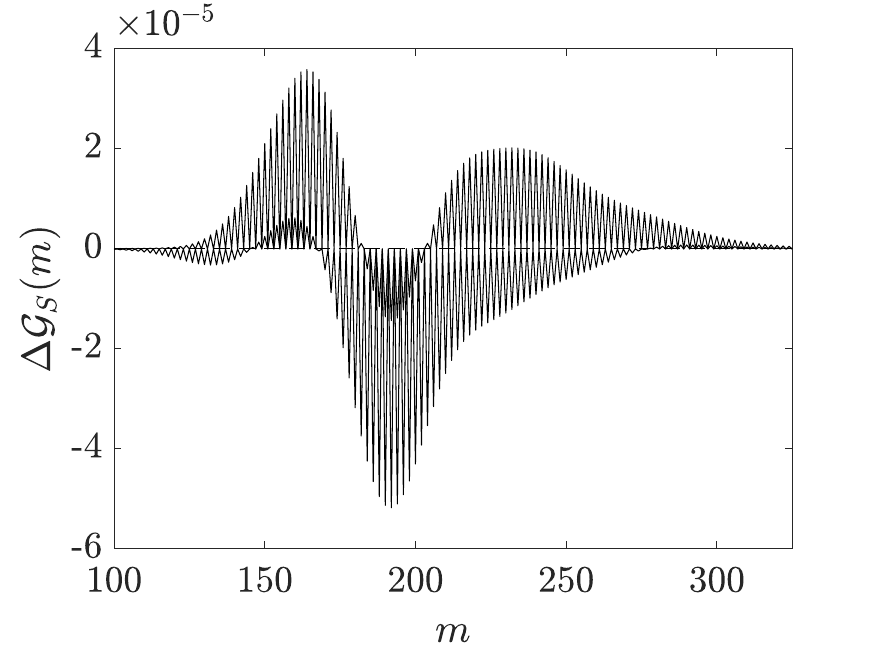}
\par\end{centering}
\caption{Upper plot: Matrix-P simulations of the total photo-count probability
(solid blue lines) versus the exact multi-mode squeezed state photon
counting distribution for GBS (dashed black lines) for squeezed states
with uniform squeezing parameter $\boldsymbol{r}=[0.75,\dots,0.75]$
and a Haar random unitary matrix of size $M=300$. Numerical probabilities
are obtained by averaging over ensembles of size $E_{S}=1.2\times10^{6}$.
Lower plot: Difference errors of the exact and matrix-P simulated
distributions, which are so small that they are not visible. \label{fig:Exact_+C_PNR_comp}}
\end{figure}

\begin{figure}
\begin{centering}
\includegraphics[width=0.4\textwidth]{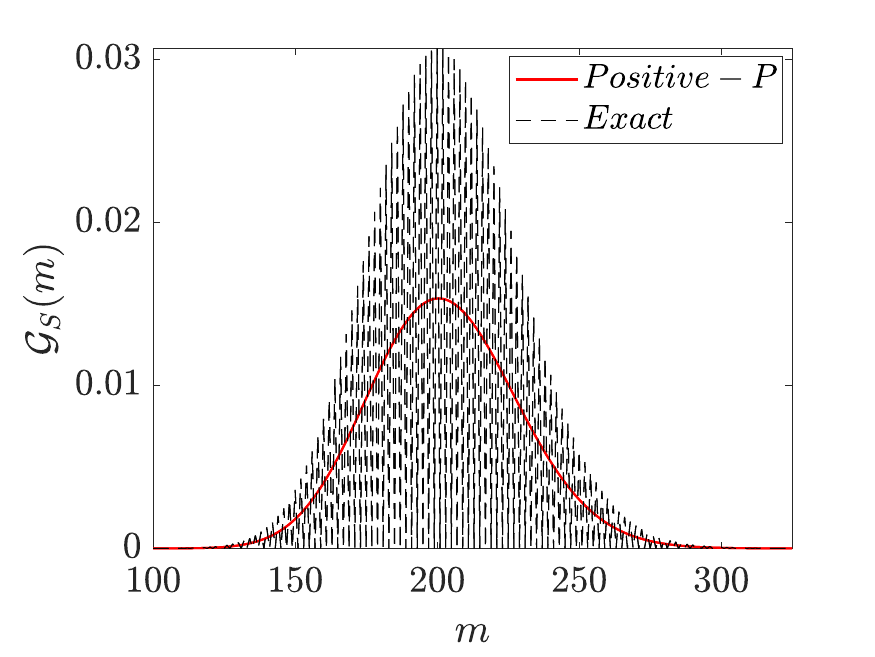}
\par\end{centering}
\begin{centering}
\includegraphics[width=0.4\textwidth]{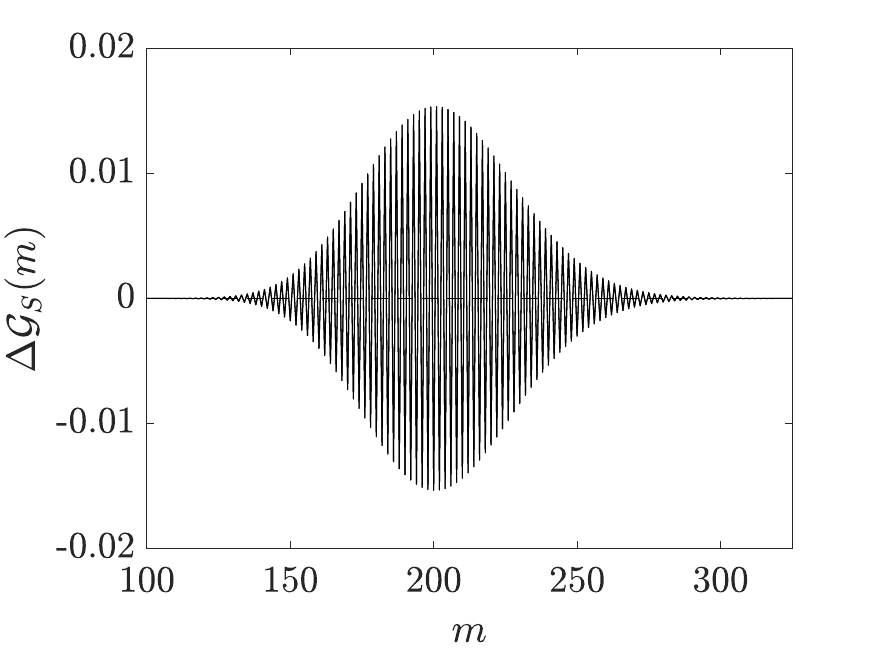}
\par\end{centering}
\caption{Description follows Fig.(\ref{fig:Exact_+C_PNR_comp}) except now
the positive-P phase-space distribution is used for simulations of
the total photo-count probability (solid red lines). Differences are
$\approx10^{3}$ times larger than matrix-P and are clearly visible.
These are caused by a skewed distribution requiring enormous sample
numbers to reach full convergence. \label{fig:Exact_+P_PNR_comp}}
\end{figure}

In this regime the matrix-P method gives far lower sampling and difference
errors than the +P method, by many orders of magnitude. This can be
seen in figures (\ref{fig:Exact_+C_PNR_comp}) and (\ref{fig:Exact_+P_PNR_comp}),
where matrix-P converges to the exact distribution while the +P simulation
is far from exact. The lack of convergence of +P in the lossless case
is due to the distribution of stochastic trajectories being extremely
skewed \citep{drummond2026matrix}, requiring an exponentially large
number of samples to resolve the even-odd oscillations. We show results
here for $M=300$ modes, in order to have adequate resolution in the
figure, but the same behavior occurs for all mode numbers. We have
simulated up to $16000$ modes using matrix-P methods, and excellent
agreement was found with exact results. The even-odd oscillations
are observed in all cases with very low sampling errors.Similar oscillations
occur in the more sensitive higher dimensional tests where detectors
are grouped into more than one subgroup.

\begin{figure}
\begin{centering}
\includegraphics[width=0.95\columnwidth]{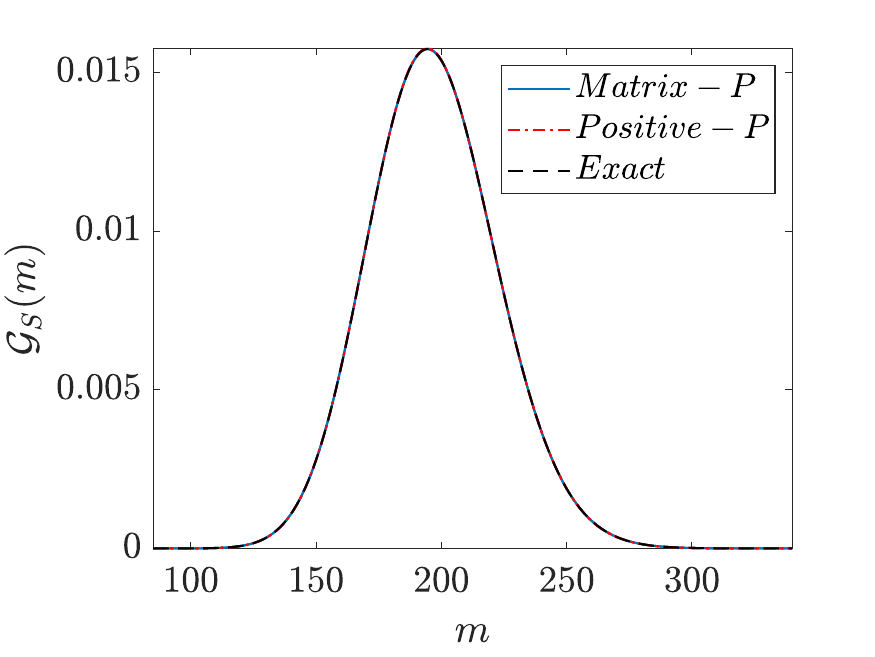}
\par\end{centering}
\begin{centering}
\includegraphics[width=0.45\columnwidth]{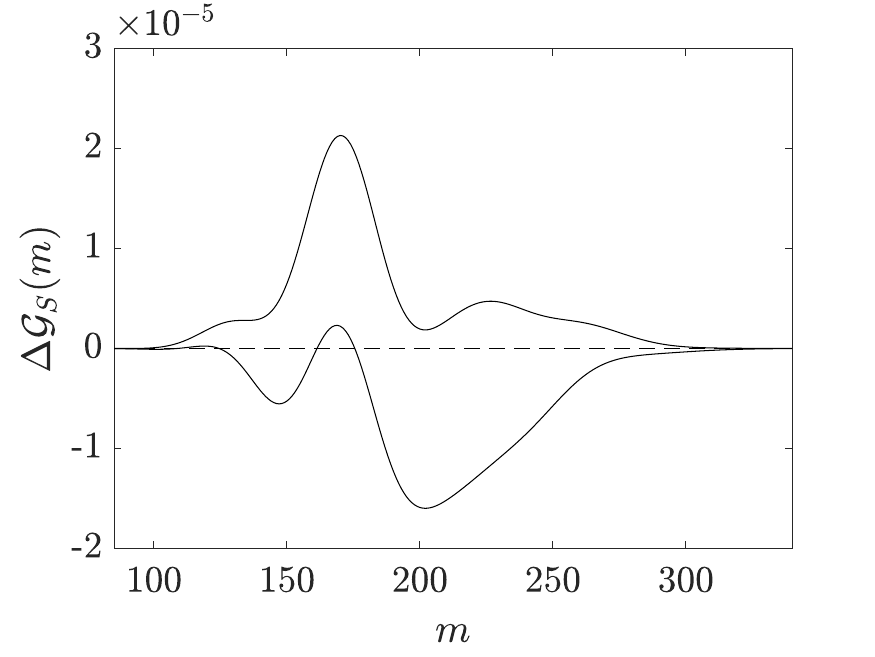}\includegraphics[width=0.45\columnwidth]{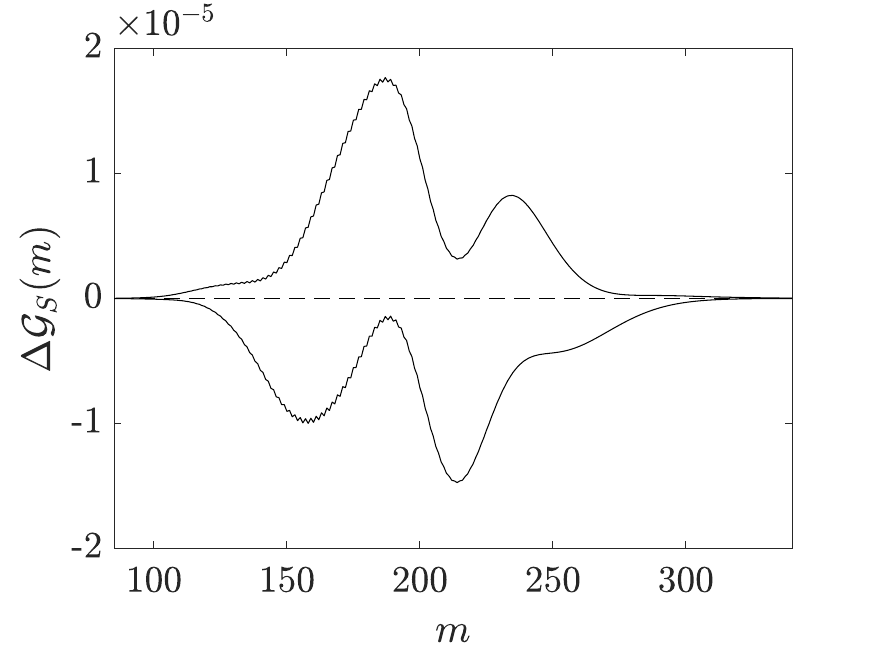}
\par\end{centering}
\caption{Upper plot: Comparison of the matrix-P (solid blue line), positive-P
(dash dot red line) and exact lossy (dashed black line) total photo-count
distribution for the network in Fig.(\ref{fig:Exact_+C_PNR_comp})
with $\approx3\%$ uniform photon loss. Both the positive-P and matrix-P
distributions converge to the exact case in this regime. Bottom plots:
Typical difference errors (showing lines for $\pm1\sigma$) of the
matrix-P (left plot) and positive-P (right plot) simulated and exact
distributions. \label{fig:Combined_lossy_totalcounts}}
\end{figure}

Current GBS with PNR detector experiments \citep{madsenQuantumComputationalAdvantage2022}
are far from lossless, and when losses are present at current levels,
the +P distribution can be used \citep{dellios2025validationPNR}.
When the parity symmetry is violated due to losses the matrix-P method
is still accurate, but its utility is diminished. In figure (\ref{fig:Combined_lossy_totalcounts})
we compare the matrix-P and +P total photo-count distributions with
the exact lossy distribution \citep{deshpandeQuantumComputationalAdvantage2022a}
for uniform loss rates of $\approx3\%$. In this case the even and
odd oscillations vanish, and both the matrix and +P distributions
converge to the exact, lossy distribution with small errors. 

While this use of the matrix-P method is not required for current
experiments, it would be essential for low-loss experiments, where
a large quantum advantage is expected. These enormous reductions in
sampling error show that the method may have applications in other
areas of many-body physics.

\section{Conclusions\label{sec:Conclusions}}

In summary, we have derived a general theory of matrix phase-space
representations. These combine phase-space and group theoretic approaches,
allowing one to include symmetries corresponding to global conservation
laws in a projected basis. Symmetries are ubiquitous in quantum theory,
and we anticipate further applications of matrix representations,
some of which will be covered in subsequent publications.

As one application, we have shown \citep{drummond2026matrix} that
for validating the exponentially hard problem of low loss gaussian
boson sampling GBS, much faster convergence is obtained for photo-count
distributions at large mode number using a matrix phase-space representation
compared with other methods. The improvements in accuracy and scalability
are because the representation projects only the physically relevant,
conserved part of Hilbert space, which is a universal issue in any
large quantum system with symmetries.

Global symmetries are found widely in quantum physics, forming the
basis of physical conservation laws \citep{Noether1918}. The methods
described are generic to all quantum systems with symmetries that
lead to conservation laws or near conservation laws. Including these
in the underlying representation of the physics is clearly advantageous.
While there are exponentially many other degrees of freedom, phase-space
methods give a scalable means of treating these, provided sampling
errors are bounded.

As a result, the examples described here are only indicative of potential
applications. Even simple parity symmetries are present in numerous
systems such as two-photon absorbers \citep{Deuar:2002,Gilchrist_Gardiner_PD_PPR_Application_Validity},
optical cat state encoding for error correction \citep{hastrup2022all,su2022universal},
and quantum metrology \citep{frerot2024symmetry}. We save a treatment
of such cases for future work. 

\subsection*{Acknowledgements}

This publication was made possible through an NTT Phi Laboratories
grant, and support of Grant 62843 from the John Templeton Foundation.
The opinions expressed in this publication are those of the author(s)
and do not necessarily reflect the views of the John Templeton Foundation.

\subsection*{Data Availability Statement}

All phase-space simulations were performed using the publicly available
software package, xQSIM \citep{Drummond2025}, which is written in
the MATLAB programming language. 

\section*{Appendix A: projection operators}

In this Appendix, we prove that the expansion in Eq.(\ref{eq:multipple symmetry projected CS})
defines a projection operator such that 
\begin{equation}
\left\Vert \boldsymbol{\alpha}\right\rangle _{\tilde{p}}=\hat{\Pi}_{\tilde{p}}\left\Vert \boldsymbol{\alpha}\right\rangle ,
\end{equation}
where $\hat{\Pi}_{\tilde{p}}$ has the usual projection operator properties:
\begin{equation}
\hat{\Pi}_{\tilde{p}}\hat{\Pi}_{\tilde{p}'}=\text{\ensuremath{\delta_{\tilde{p}\tilde{p}'}\hat{\Pi}_{\tilde{p}}}.}\label{eq:projection operator orthogonality}
\end{equation}

\subsection*{Symmetry projectors}

To prove that $\hat{\Pi}_{\tilde{p}}$ defines a projection operator
on bosonic Hilbert space, we initially treat a single group transformation
$\bm{U}^{q}$, then multiple groups, $\bm{U}^{\left(\tilde{q}\right)}=\bm{U}_{1}^{q_{1}}\bm{U}_{2}^{q_{2}}\cdots\bm{U}_{\mathcal{N}}^{q_{\mathcal{N}}}$. 

The coherent states $\left\Vert \boldsymbol{\alpha}\right\rangle $
are a complete basis, hence a definition of a projection operator
$\hat{\Pi}_{p}$ is obtained through defining its action on an arbitrary
coherent state, which in this case is:
\begin{align}
\hat{\Pi}_{p}\left\Vert \boldsymbol{\alpha}\right\rangle  & \equiv\frac{1}{\mathcal{\mathcal{M}}}\left[\sum_{q=0}^{\mathcal{\mathcal{M}}-1}e^{i\phi q\left(\hat{g}-p\right)}\right]\left\Vert \boldsymbol{\alpha}\right\rangle .\\
 & =\frac{1}{\mathcal{\mathcal{M}}}\sum_{q=0}^{\mathcal{\mathcal{M}}-1}e^{-ipq\phi}\left\Vert \bm{U}^{q}\boldsymbol{\alpha}\right\rangle 
\end{align}
For a product of two projectors of this type, one obtains:
\begin{align}
\hat{\Pi}_{p}\left\Vert \boldsymbol{\alpha}\right\rangle _{p'} & \equiv\hat{\Pi}_{p}\hat{\Pi}_{p'}\left\Vert \boldsymbol{\alpha}\right\rangle \nonumber \\
 & =\frac{\hat{\Pi}_{p}}{\mathcal{\mathcal{M}}}\sum_{q'=0}^{\mathcal{\mathcal{M}}-1}e^{-ip'q'\phi}\left\Vert \bm{U}^{q'}\boldsymbol{\alpha}\right\rangle .
\end{align}
Using the composition property of the unitary operators, this becomes
\begin{equation}
\hat{\Pi}_{p}\left\Vert \boldsymbol{\alpha}\right\rangle _{p'}=\frac{1}{\mathcal{\mathcal{M}}^{2}}\sum_{q,q'=0}^{\mathcal{\mathcal{M}}-1}e^{-i\left(pq+p'q'\right)\phi}\left\Vert \bm{U}^{q+q'}\boldsymbol{\alpha}\right\rangle .
\end{equation}
Since $q'$ is a dummy variable, its range can be changed by numbers
of modulo $\mathcal{M}$ (recall the $\mathcal{M}$-th root of unity
property Eq.(\ref{eq:cyclic property-1})) 
\begin{align}
\hat{\Pi}_{p}\left\Vert \boldsymbol{\alpha}\right\rangle _{p'} & =\frac{1}{\mathcal{\mathcal{M}}^{2}}\sum_{q=0}^{\mathcal{\mathcal{M}}-1}\sum_{q'=-q}^{\mathcal{\mathcal{M}}-1-q}e^{-i\left(pq+p'q'\right)\phi}\left\Vert \bm{U}^{q+q'}\boldsymbol{\alpha}\right\rangle .
\end{align}

Upon defining $q_{+}=q+q'$ and $\delta=p-p'$, where $q_{+}\in\left[0,\mathcal{\mathcal{\mathcal{M}}}-1\right]$,
since $p'q'=\left(p-\delta\right)\left(q_{+}-q\right)$, the sums
are rewritten as:
\begin{align}
\hat{\Pi}_{p}\left\Vert \boldsymbol{\alpha}\right\rangle _{p'} & =\frac{1}{\mathcal{\mathcal{M}}^{2}}\sum_{q=0}^{\mathcal{\mathcal{M}}-1}e^{-i\delta q\phi}\left(\sum_{q_{+}=0}^{\mathcal{\mathcal{M}}-1}e^{-i\left(p-\delta\right)q_{+}\phi}\left\Vert \bm{U}^{q_{+}}\boldsymbol{\alpha}\right\rangle \right).
\end{align}
From properties of the discrete Fourier transform, the first summation
is a delta-function, therefore:
\begin{align}
\hat{\Pi}_{p}\left\Vert \boldsymbol{\alpha}\right\rangle _{p'} & =\frac{\delta_{pp'}}{\mathcal{\mathcal{M}}}\left(\sum_{q_{+}=0}^{\mathcal{\mathcal{M}}-1}e^{-ipq_{+}\phi}\left\Vert \bm{U}^{q_{+}}\boldsymbol{\alpha}\right\rangle \right)=\delta_{pp'}\left\Vert \boldsymbol{\alpha}\right\rangle _{p},
\end{align}
proving that 
\begin{equation}
\hat{\Pi}_{p}\hat{\Pi}_{p'}=\delta_{pp'}\hat{\Pi}_{p}.
\end{equation}

Hence the operator $\hat{\Pi}_{p}$ satisfies the orthogonal law of
projection operators. It is also hermitian since its only eigenvalues
are $1$ or $0$, which are real.

\subsection*{Composite projectors}

Consider a composite group transformation as in Eq. (\ref{eq:projector_composite}),
where we recall
\begin{equation}
\hat{\Pi}_{\tilde{p}}\left\Vert \boldsymbol{\alpha}\right\rangle \equiv\frac{1}{\mathcal{\mathcal{M}}}\sum_{\tilde{q}}e^{-i\tilde{p}\tilde{q}\tilde{\phi}}\left\Vert \bm{U}^{\left(\tilde{q}\right)}\boldsymbol{\alpha}\right\rangle .
\end{equation}

As above, we now prove the composite projectors satisfy the orthogonal
relation. Applying the projection twice, combined with the fact that
the unitary matrices commute, gives:
\begin{align}
\hat{\Pi}_{\tilde{p}}\left\Vert \boldsymbol{\alpha}\right\rangle _{\tilde{p}'} & =\frac{1}{\mathcal{\mathcal{M}}^{2}}\sum_{\tilde{q},\tilde{q}'}e^{-i\left(\tilde{p}\tilde{q}+\tilde{p}'\tilde{q}'\right)\tilde{\phi}}\left\Vert \bm{U}^{\left(\tilde{q}+\tilde{q}'\right)}\boldsymbol{\alpha}\right\rangle .\label{eq:composite-projector-product}
\end{align}
 Exploiting the property Eq.(\ref{eq:cyclic property-1}), and following
the single group procedure, one similarly obtains 
\begin{equation}
\hat{\Pi}_{\tilde{p}}\hat{\Pi}_{\tilde{p}'}=\delta_{\tilde{p}\tilde{p}'}\hat{\Pi}_{\tilde{p}},\label{eq:product_of_projectors}
\end{equation}
proving the composite operator satisfies the orthogonal relation of
projection operators.

\subsection*{Completeness}

To prove that the set of projectors is complete, we use properties
of a discrete Fourier transform such that for a single symmetry group:
\begin{align}
\sum_{p=0}^{\mathcal{\mathcal{M}}-1}\hat{\Pi}_{p}\left\Vert \boldsymbol{\alpha}\right\rangle  & =\frac{1}{\mathcal{\mathcal{M}}}\sum_{q=0}^{\mathcal{\mathcal{M}}-1}\left[\sum_{p=0}^{\mathcal{\mathcal{M}}-1}e^{-ipq\phi}\right]\left\Vert \bm{U}^{q}\boldsymbol{\alpha}\right\rangle \nonumber \\
 & =\sum_{q=0}^{\mathcal{\mathcal{M}}-1}\delta_{q0}\left\Vert \bm{U}^{q}\boldsymbol{\alpha}\right\rangle .
\end{align}

Since $\left\Vert \bm{U}^{0}\boldsymbol{\alpha}\right\rangle =\left\Vert \boldsymbol{\alpha}\right\rangle $,
this result shows that the sum over all projectors of a given type
is the unit operator, proving completeness: 
\begin{equation}
\sum_{p=0}^{\mathcal{\mathcal{M}}-1}\hat{\Pi}_{p}=\hat{1}.
\end{equation}
An analogous result, namely: 
\begin{equation}
\sum_{\tilde{p}}\hat{\Pi}_{\tilde{p}}=\hat{1}\label{eq:projector completeness}
\end{equation}
follows for composite projections, using the steps given above recursively
for each type of projection.

In summary, we have shown that defining $\hat{\Pi}_{\tilde{p}}$ and
$\left\Vert \boldsymbol{\alpha}\right\rangle _{\tilde{p}}$ using
quantum superpositions of coherent states corresponds to a projection
operator and a projected coherent state respectively. 

\section*{Appendix B: symmetry examples}

We now give examples of symmetries, and how their unitary matrices
are generated, to clarify results in the main text. 

One purpose of these examples is to show that, due to the over-completeness
of the coherent states, there are other choices of mappings in addition
to those in the existence theorem. These can be chosen to be more
compact than given by the existence theorem. While these examples
are simple, they are closely related to many current quantum technologies. 

\subsection{Phase symmetry}

The simplest example, an application of which is presented in the
main text, is a phase symmetry. In this case the group generator,
and its eigenvalues, are $\bm{g}=\boldsymbol{h}=\bm{I}$ such that
\begin{equation}
\bm{U}^{q}=\bm{I}e^{iq\phi}.\label{eq:phase-symmetry}
\end{equation}
Hence the transformed coherent amplitudes are phase shifted by $q\phi$,
so that $\bm{U}^{q}\boldsymbol{\alpha}=\boldsymbol{\alpha}\exp\left(iq\phi\right)$. 

In the limit $\mathcal{M}\rightarrow\infty$, a phase symmetry physically
corresponds to a total number conservation law. For all values of
$\mathcal{M}$, the generating operator is the number operator.

\subsubsection{Parity symmetry}

The simplest nontrivial phase symmetry case is an $\mathcal{M}=2$
parity symmetry. If we choose an $M=2$ element group representation,
then $\phi=\pi$ and the unitary is expanded as 
\begin{equation}
\bm{U}^{q}=\boldsymbol{I}\cos\left(\pi q\right)+i\boldsymbol{I}\sin\left(\pi q\right),
\end{equation}
such that $\bm{U}^{0}\boldsymbol{\alpha}=\boldsymbol{\alpha}$ and
$\bm{U}^{1}\boldsymbol{\alpha}=-\boldsymbol{\alpha}$, i.e. a reflection
in the space of coherent amplitudes. 

For this case, the projected states are 
\begin{equation}
\left\Vert \boldsymbol{\alpha}\right\rangle _{p}=\frac{1}{2}\left(\left\Vert \boldsymbol{\alpha}\right\rangle +\left(-1\right)^{p}\left\Vert -\boldsymbol{\alpha}\right\rangle \right).
\end{equation}

Including normalization factors, for the single mode case one has
a Schrödinger cat state \citep{yurke1986generating}, with $\left|\psi_{s}\right\rangle \equiv\left(\left\Vert x\right\rangle +\left\Vert -x\right\rangle \right)/\sqrt{4C}$,
given an amplitude $x$, so that:
\begin{align}
\hat{\rho} & =\frac{1}{4C}\left[\left\Vert x\right\rangle +\left\Vert -x\right\rangle \right]\left[\left\langle x\right\Vert +\left\langle -x\right\Vert \right],\label{eq:Schrodinger-cat}
\end{align}
where $C=\cosh\left(\left|x\right|^{2}\right)$.

This is the symmetry that is used to study gaussian boson sampling
(GBS) in the lossless and low-loss limit. It is applicable because
the input states used are squeezed states which only generate even
parity pairs of photons, and the parity doesn't change if there are
no losses. If there are losses, these can be readily treated by adding
additional reservoir modes to absorb all the lost photons. 

\subsubsection{Number conservation}

If we choose $\mathcal{\mathcal{M}}>2$, then one obtains a discrete
phase rotation in the complex space of a single coherent amplitude.
For example, $\mathcal{M}=4$ corresponds to a $\phi=\pi/2$ rotation
with projected states 
\begin{equation}
\left\Vert \boldsymbol{\alpha}\right\rangle _{p}=\frac{1}{4}\left(\left\Vert \boldsymbol{\alpha}\right\rangle +(-i)^{p}\left\Vert i\boldsymbol{\alpha}\right\rangle +\left(-1\right)^{p}\left\Vert -\boldsymbol{\alpha}\right\rangle +(i)^{p}\left\Vert -i\boldsymbol{\alpha}\right\rangle \right).
\end{equation}

Number conservation laws correspond to the limit of $\mathcal{M\rightarrow\infty}$.
This doesn't change the matrix $\bm{g}=\bm{I}$, but it does change
the normalization used, as shown in Subsection \ref{subsec:Inner-products-phase-sym}.
From the property Eq.(\ref{eq:continuum limit eigenvalue}), one can
show that the single-mode projected states are the number states:
\begin{equation}
\lim_{\mathcal{M}\rightarrow\infty}\hat{g}\left\Vert \alpha\right\rangle _{p}=\sum_{n=0}^{\infty}\frac{\alpha^{n}}{\sqrt{n!}}\left|n\right\rangle .
\end{equation}

\subsection{Translational symmetry}

The next example is translational symmetry. Following the definitions
Eq.(\ref{eq:diagonal-h-matrix-1}) and Eq.(\ref{eq:diagonal-U-1}),
if we let the coherent mode index be a lattice point in one space
dimension, then the eigenvalues correspond to discrete momenta with
$\tilde{h}_{k}=0,1,2,..\mathcal{M}-1$, and the $\bm{V}$matrix is
a discrete Fourier transform unitary with:
\begin{equation}
V_{kj}=\frac{1}{\sqrt{\mathcal{M}}}e^{-i\phi kj}.
\end{equation}
This corresponds physically to a spatial translation symmetry on a
torus, and hence to a momentum conservation law. 

Unlike the phase symmetry example, for a translation symmetry we have
$M=\mathcal{M}$, as the number of eigenvalues corresponds to the
number of spatial lattice points. 

\subsubsection{Double-well case}

In the $M=2$ case, one has a spatial lattice with two lattice sites,
i.e. a double well. Choosing $\mathcal{M}=2$, the diagonalizing matrix
is:
\begin{equation}
\bm{V}=\frac{1}{\sqrt{2}}\left[\begin{array}{cc}
1 & 1\\
1 & -1
\end{array}\right].
\end{equation}

In this case $\boldsymbol{h}=diag\left[0,1\right]$, and the symmetry
generator matrix $\bm{g}$ is:
\begin{equation}
\bm{g}=\frac{1}{2}\left[\begin{array}{cc}
1 & -1\\
-1 & 1
\end{array}\right].
\end{equation}
Since $\phi=\pi$, the general unitary is expanded as 
\begin{equation}
\boldsymbol{U}^{q}=\boldsymbol{\sigma}^{x}+\boldsymbol{g}+\left(\cos\left(q\pi\right)+i\sin\left(\pi q\right)\right)\boldsymbol{g},
\end{equation}
with $\bm{\sigma}^{x}$ being the Pauli $x$-spin matrix. 

Defining $\boldsymbol{\alpha}'\equiv\left[\begin{array}{c}
\alpha_{2}\\
\alpha_{1}
\end{array}\right]$, the projected states are:
\begin{equation}
\left\Vert \boldsymbol{\alpha}\right\rangle _{p}=\frac{1}{2}\left(\left\Vert \boldsymbol{\alpha}\right\rangle +\left(-1\right)^{p}\left\Vert \boldsymbol{\alpha}'\right\rangle \right),
\end{equation}
where in each case the second term corresponds to a swap of the two
coherent amplitudes.

\subsection{Combined translational and phase symmetry}

\subsubsection{Parity symmetric double-well}

As an example with two symmetries, we take a case of combined parity
and double-well translational symmetry, so that:
\begin{equation}
\boldsymbol{U}^{(\tilde{q})}=\bm{U}_{1}^{q_{1}}\bm{U}_{2}^{q_{2}}=e^{i\pi q_{1}}e^{i\pi q_{2}\boldsymbol{g}_{2}}.
\end{equation}

Each unitary can be applied independently, leading to more independent
projected states in the superposition. Therefore, there are four discrete
projected states in the basis, and one has: 
\begin{align}
\left|\boldsymbol{\alpha}\right\rangle _{00} & =\frac{1}{4}\left(\left|\boldsymbol{\alpha}\right\rangle +\left|\boldsymbol{\alpha}'\right\rangle +\left|-\boldsymbol{\alpha}\right\rangle +\left|-\boldsymbol{\alpha}'\right\rangle \right)\nonumber \\
\left|\boldsymbol{\alpha}\right\rangle _{10} & =\frac{1}{4}\left(\left|\boldsymbol{\alpha}\right\rangle +\left|\boldsymbol{\alpha}'\right\rangle -\left|-\boldsymbol{\alpha}\right\rangle -\left|-\boldsymbol{\alpha}'\right\rangle \right)\nonumber \\
\left|\boldsymbol{\alpha}\right\rangle _{01} & =\frac{1}{4}\left(\left|\boldsymbol{\alpha}\right\rangle -\left|\boldsymbol{\alpha}'\right\rangle +\left|-\boldsymbol{\alpha}\right\rangle -\left|-\boldsymbol{\alpha}'\right\rangle \right)\nonumber \\
\left|\boldsymbol{\alpha}\right\rangle _{11} & =\frac{1}{4}\left(\left|\boldsymbol{\alpha}\right\rangle -\left|\boldsymbol{\alpha}'\right\rangle -\left|-\boldsymbol{\alpha}\right\rangle +\left|-\boldsymbol{\alpha}'\right\rangle \right).
\end{align}

\subsubsection{Extended parity symmetry}

Next, take a Schrödinger cat state that knows neither where it is
nor if it is dead or alive, given by:
\begin{equation}
\left|\psi_{t}\right\rangle =\frac{\sum_{j=1}^{M}\left(\left\Vert x\right\rangle ^{j}+\left\Vert -x\right\rangle ^{j}\right)\prod_{k\neq j}\left\Vert 0\right\rangle ^{k}}{\sqrt{4M\left(\cosh\left(x^{2}\right)+M-1\right)}}.
\end{equation}
where $\left\Vert x\right\rangle ^{j}\equiv\exp\left(x_{j}a_{j}^{\dagger}\right)\left|0\right\rangle $
is a coherent state localized at spatial mode $j$. 

Using number states to carry out calculations would require an exponentially
large basis, since with a number cutoff at $N_{c}$, the basis state
dimension grows as $\exp\left(M\ln N_{c}\right)$. Just as with the
single-mode cat state, the Glauber P-function is singular, while the
Wigner distribution is oscillatory. In this case the Q-function is
given by a sum of $4M^{2}$ terms, each in 2M-dimensions, and has
a growing sampling error with mode number $M$. 

The +P-function is a sum of delta-function terms that grows polynomially
as $4M^{2}$, which can be sampled with reasonable efficiency, but
it does not take advantage of any of the symmetries involved. 

By comparison, the matrix-P representation allows one to represent
this as a single delta-function in the extended phase-space, which
implies that the translational cat state is a projected coherent state
with indices $\left(p,q\right)=\left(0,0\right)$: 
\begin{equation}
\left|\psi_{t}\right\rangle \propto\left\Vert x\right\rangle _{00}.
\end{equation}
As a result, one can write a vector $\bm{x}=\left(x,0,...\right)$
for the coherent amplitude in the first lattice location, then obtaining
a matrix-P distribution as a single delta-function:
\begin{equation}
P\left(\vec{\lambda},t\right)=\delta(\bm{\alpha}-\bm{x})\delta(\bm{\beta}-\bm{x})\delta(\underline{\Omega}-\underline{\Omega}^{(0)}),
\end{equation}
where
\begin{equation}
\Omega_{\tilde{p}\tilde{q}}^{(0)}\equiv\delta_{\tilde{p}}\delta_{\tilde{q}}.
\end{equation}

The gauge matrices and projectors are $2\mathcal{M}\times2\mathcal{M}$
dimensional, but the translational cat-state requires only a single
non-zero entry. This is a more compact description than results from
using the existence theorem given above.

\section*{Appendix C: Matrix-P identities}

In this Appendix we present more detailed calculations of the matrix-P
operator identities, following the same general structure as the main
text. That is, first we treat operators acting on state kets, then
the kernel operator matrix. 

For notational simplicity, we assume $\hat{a}_{j}$ is the operator
basis where all unitary transformations have diagonal symmetry group
generators $\bm{g}_{\nu}$, with a diagonal vector of integer quantum
numbers $h_{j\nu}$, so that:
\begin{align}
U_{jk}^{\left(\tilde{q}\right)} & =\delta_{jk}\exp\left(i\sum_{\nu=1}^{\mathcal{N}}\phi_{\nu}q_{\nu}h_{j\nu}\right)\equiv\delta_{jk}\exp\left(i\tilde{\phi}\tilde{q}\tilde{h}_{j}\right),
\end{align}
where for compactness we use an implicit sum over group labels $\nu$
for triple products of $\mathcal{N}$-vectors.

\subsection*{State ket identities for $\hat{a}$}

Starting with the multi-symmetry projected coherent state Eq.(\ref{eq:multipple symmetry projected CS}),
one has 
\begin{equation}
\hat{a}_{k}\left\Vert \boldsymbol{\alpha}\right\rangle _{\tilde{p}}=\frac{1}{\mathcal{\mathcal{M}}}\sum_{\tilde{q}}e^{i\tilde{\phi}\tilde{q}\left(\tilde{h}_{k}-\tilde{p}\right)}\alpha_{k}\left\Vert \bm{U}^{(\tilde{q})}\boldsymbol{\alpha}\right\rangle .
\end{equation}
where we employ the relation that $\hat{a}_{j}\left\Vert \bm{U}^{(\tilde{q})}\boldsymbol{\alpha}\right\rangle =U_{jk}^{\left(\tilde{q}\right)}\alpha_{k}\left\Vert \bm{U}^{(\tilde{q})}\boldsymbol{\alpha}\right\rangle $.
Using the inverse transform Eq.(\ref{eq:Inverse transform}), gives
the result that:
\begin{align}
\hat{a}_{k}\left\Vert \boldsymbol{\alpha}\right\rangle _{\tilde{p}} & =\frac{\alpha_{k}}{\mathcal{\mathcal{M}}}\sum_{\tilde{q}=0}^{\mathcal{M}-1}\sum_{\tilde{p}'=0}^{\mathcal{M}-1}e^{i\left(\tilde{h}_{k}+\tilde{p}'-\tilde{p}\right)\tilde{q}\tilde{\phi}}\left\Vert \boldsymbol{\alpha}\right\rangle _{\tilde{p}'}.
\end{align}
We define the Fourier transform above as a delta function:
\begin{align}
\sigma_{k}^{\tilde{p}\tilde{p}'} & \equiv\frac{1}{\mathcal{\mathcal{M}}}\sum_{\tilde{q}=0}^{\mathcal{M}-1}e^{i\left(\tilde{h}_{k}+\tilde{p}'-\tilde{p}\right)\tilde{q}\tilde{\phi}}=\delta_{\tilde{p}-\tilde{h}_{k},\tilde{p}'}^{(\mathcal{M})}\label{eq:General_sigma_matrix}
\end{align}
which is  the periodic Kronecker delta function Eq.(\ref{eq:periodic_kronecker}).
Hence the matrix-P annihilation operator ket identity is 
\begin{align}
\hat{a}_{k}\left\Vert \boldsymbol{\alpha}\right\rangle _{\tilde{p}} & =\sigma_{k}^{\tilde{p}\tilde{p}'}\alpha_{k}\left\Vert \boldsymbol{\alpha}\right\rangle _{\tilde{p}'}=\alpha_{k}\left\Vert \boldsymbol{\alpha}\right\rangle _{\tilde{p}-\tilde{h}_{k}},
\end{align}
with an Einstein summation convention for repeated $\tilde{p}$ indices. 

The physical explanation is very simple. If a particle in mode $k$
is removed from the state, its quantum numbers are a vector $\tilde{h}_{k}$
which is subtracted from the total quantum numbers $\tilde{p}$, as
one would intuitively expect. If there is more than one symmetry group,
so $\mathcal{N}>1$, there is a vector of quantum numbers, otherwise
it is a scalar.

For the case of translational symmetry, removing a particle with momentum
$h_{k}$ reduces the total momentum by $h_{k}$. 

\subsubsection*{Phase symmetry case:}

In the $\mathcal{N}=1$, $\mathcal{M}=2$ phase symmetry case, $h_{k}=1$
and this simplifies to give:
\begin{align}
\hat{a}_{j}\left\Vert \boldsymbol{\alpha}\right\rangle _{p} & =\alpha_{j}\left\Vert \boldsymbol{\alpha}\right\rangle _{p-1}.
\end{align}

For this example the index $p$ is a scalar equal to the total particle
number. The annihilation operator decreases the total number by one,
and multiplies the state by the coherent amplitude $\alpha_{j}$. 

\subsection*{State ket identities for $\hat{a}^{\dagger}$}

Next, we derive a ket identity for $\hat{a}^{\dagger}$, which requires
using the expansion
\begin{equation}
\left\Vert \bm{U}^{(\tilde{q})}\boldsymbol{\alpha}\right\rangle =\exp\left(\bm{U}^{(\tilde{q})}\boldsymbol{\alpha}\cdot\bm{a}^{\dagger}\right)\left|0\right\rangle ,
\end{equation}
such that from Eq.(\ref{eq:multipple symmetry projected CS})
\begin{equation}
\hat{a}_{j}^{\dagger}\left\Vert \boldsymbol{\alpha}\right\rangle _{\tilde{p}}=\hat{a}_{j}^{\dagger}\frac{1}{\mathcal{\mathcal{M}}}\sum_{\tilde{q}=0}^{\mathcal{\mathcal{M}}-1}e^{-i\tilde{p}\tilde{q}\phi}\exp\left(\left(\bm{U}^{(\tilde{q})}\boldsymbol{\alpha}\right)_{k}\cdot\bm{a}^{\dagger}\right)\left|0\right\rangle ,
\end{equation}
where:
\begin{equation}
\left(\bm{U}^{\left(\tilde{q}\right)}\boldsymbol{\alpha}\right)_{k}=e^{i\tilde{h}_{k}\tilde{q}\tilde{\phi}}\alpha_{k}.
\end{equation}

From the chain rule for variable changes, and the inverse discrete
Fourier transform,
\begin{align}
\hat{a}_{k}^{\dagger}\left\Vert \boldsymbol{\alpha}\right\rangle _{\tilde{p}} & =\frac{1}{\mathcal{\mathcal{M}}}\sum_{\tilde{q}=0}^{\mathcal{M}-1}e^{-i\left(\tilde{h}_{k}+\tilde{p}\right)\cdot\tilde{q}\tilde{\phi}}\frac{\partial}{\partial\alpha_{k}}\sum_{\tilde{p}'=0}^{\mathcal{M}-1}e^{i\tilde{p}'\tilde{q}\tilde{\phi}}\left\Vert \boldsymbol{\alpha}\right\rangle _{\tilde{p}'}.
\end{align}
Upon defining 
\begin{align}
\bar{\sigma}_{k}^{\tilde{p}\tilde{p}'} & \equiv\frac{1}{\mathcal{\mathcal{M}}}\sum_{\tilde{q}=0}^{\mathcal{M}-1}e^{i\left(\tilde{p}'-\tilde{p}-\tilde{h}_{k}\right)\tilde{q}\tilde{\phi}}=\delta_{\tilde{p}+\tilde{h}_{k},\tilde{p}'}^{(\mathcal{M})},
\end{align}
we obtain the following identity
\begin{align}
\hat{a}_{k}^{\dagger}\left\Vert \boldsymbol{\alpha}\right\rangle _{\tilde{p}} & =\bar{\sigma}_{k}^{\tilde{p}'\tilde{p}}\partial_{k}\left\Vert \boldsymbol{\alpha}\right\rangle _{\tilde{p}'}=\partial_{k}\left\Vert \boldsymbol{\alpha}\right\rangle _{\tilde{p}+\tilde{h}_{k}}.
\end{align}

Hence, when a particle in mode $k$ is added to the state, the corresponding
projected eigenvalues or quantum numbers $\tilde{h}_{k}$ are added,
as one would expect.

\subsubsection*{Phase symmetry case:}

Again, we treat a $\mathcal{N}=1$, $\mathcal{M}=2$ phase symmetry
example, where $h_{k}=1$ gives:
\begin{align}
\hat{a}_{j}^{\dagger}\left\Vert \boldsymbol{\alpha}\right\rangle _{p} & =\partial_{j}\left\Vert \boldsymbol{\alpha}\right\rangle _{p+1}.
\end{align}

Hence the creation operator differentiates the coherent state by $\alpha_{j}$,
and increases the photon number by one.

\subsection*{Vector ket identities:}

In vector notation, the transforms $\sigma_{k}^{\tilde{p}\tilde{p}'}$
and $\bar{\sigma}_{k}^{\tilde{p}\tilde{p}'}$ are the matrices $\underline{\sigma}_{k}$
and $\underline{\sigma}_{k}^{T}$, which is the transpose of $\underline{\sigma}_{k}$
in the projection indices $\tilde{p}$. Therefore, the matrices are
cyclic delta functions in each vector component, such that 
\begin{align}
\hat{a}_{k}\underline{\left\Vert \boldsymbol{\alpha}\right\rangle } & =\alpha_{k}\underline{\sigma}_{k}\underline{\left\Vert \boldsymbol{\alpha}\right\rangle }\nonumber \\
\hat{a}_{k}^{\dagger}\underline{\left\Vert \boldsymbol{\alpha}\right\rangle } & =\partial_{k}\underline{\sigma}_{k}^{T}\underline{\left\Vert \boldsymbol{\alpha}\right\rangle }.\label{eq:ket-identities}
\end{align}
As stated in the main text, we indicate matrices and vectors in the
space of projection indices with an underline, which is a different
vector space to the phase-space of mode amplitudes. 

Operator products have similar identities, which are given below:
\begin{align}
\hat{a}_{i}^{\dagger}\hat{a}_{j}\underline{\left\Vert \boldsymbol{\alpha}\right\rangle } & =\alpha_{j}\partial_{i}\underline{\sigma}_{i}^{T}\underline{\sigma}_{j}\underline{\left\Vert \boldsymbol{\alpha}\right\rangle }\nonumber \\
\hat{a}_{i}\hat{a}_{j}\underline{\left\Vert \boldsymbol{\alpha}\right\rangle } & =\alpha_{i}\alpha_{j}\underline{\sigma}_{i}\underline{\sigma}_{j}\underline{\left\Vert \boldsymbol{\alpha}\right\rangle }\nonumber \\
\hat{a}_{i}^{\dagger}\hat{a}_{j}^{\dagger}\underline{\left\Vert \boldsymbol{\alpha}\right\rangle } & =\partial_{i}\partial_{j}\underline{\sigma}_{i}^{T}\underline{\sigma}_{j}^{T}\underline{\left\Vert \boldsymbol{\alpha}\right\rangle }
\end{align}
In the parity symmetry case with $\mathcal{\mathcal{\mathcal{M}}}=2$,
the $\underline{\sigma}$ matrix is symmetric, real, and equal to
a Pauli matrix: $\underline{\sigma}=\underline{\sigma}^{T}=\underline{\sigma}^{x}$.

\subsection*{Normally ordered kernel matrix identities:}

Using the above ket identities, we now calculate the differential
identities for the matrix kernel $\hat{\underline{\Lambda}}(\vec{\alpha})$
for normal ordering, with annihilation operators to the left or creation
operators to the right. 

Using Einstein summation conventions, the left annihilation operator
identity is:
\begin{align}
\hat{a}_{k}\hat{\Lambda}_{\tilde{p}\tilde{q}} & =\alpha_{k}\sigma_{k}^{\tilde{p}\tilde{p}'}\left\Vert \boldsymbol{\alpha}\right\rangle _{\tilde{p}'}\left\langle \boldsymbol{\beta}^{*}\right\Vert _{\tilde{q}}e^{-w_{\tilde{p}\tilde{q}}\left(\vec{\alpha}\right)}\nonumber \\
 & =\alpha_{k}\sigma_{k}^{\tilde{p}\tilde{p}'}\left\Vert \boldsymbol{\alpha}\right\rangle _{\tilde{p}'}\left\langle \boldsymbol{\beta}^{*}\right\Vert _{\tilde{q}}e^{-w_{\tilde{p}\tilde{q}}\left(\vec{\alpha}\right)}e^{-w_{\tilde{p}'\tilde{q}}\left(\vec{\alpha}\right)}e^{w_{\tilde{p}'\tilde{q}}\left(\vec{\alpha}\right)}\nonumber \\
 & =\alpha_{k}\Sigma_{k}^{\tilde{p}\tilde{p}'}\left(\vec{\alpha}\right)\hat{\Lambda}_{\tilde{p}'\tilde{q}}.
\end{align}
Here we introduce the $\Sigma$ matrix, which equals the cyclic delta
function $\sigma_{k}^{\tilde{p}\tilde{p}'}$ matrix with an additional
weighting term: 
\begin{align}
\Sigma_{k}^{\tilde{p}\tilde{p}'}\left(\vec{\alpha}\right) & \equiv\sigma_{k}^{\tilde{p}\tilde{p}'}e^{w_{\tilde{p}'}\left(\vec{\alpha}\right)-w_{\tilde{p}}\left(\vec{\alpha}\right)}\nonumber \\
 & =\sigma_{k}^{\tilde{p}\tilde{p}'}\sqrt{G_{\tilde{p}'}\left(\vec{\alpha}\right)/G_{\tilde{p}}\left(\vec{\alpha}\right)}.
\end{align}
where we have used the factorization Eq.(\ref{eq:weight_factorization}).
The additional weight term is required to renormalize the kernel operator,
since one now has a $\tilde{p}'$ quantum number index. 

Similarly, the right creation operator identity is:
\begin{align}
\hat{\Lambda}_{\tilde{p}\tilde{q}}\hat{a}_{k}^{\dagger} & =\beta_{k}\left\Vert \boldsymbol{\alpha}\right\rangle _{\tilde{p}}\left\langle \boldsymbol{\beta}^{*}\right\Vert _{\tilde{q}'}\sigma_{k}^{\tilde{q}\tilde{q}'}e^{-w_{\tilde{p}\tilde{q}}}\nonumber \\
 & =\beta_{k}\Sigma_{k}^{\tilde{q}\tilde{q}'}\left(\vec{\alpha}'\right)\hat{\Lambda}_{\tilde{p}\tilde{q}'}.
\end{align}

To obtain the full kernel matrix identities we note that, assuming
the factorization Eq.(\ref{eq:weight_factorization}), the $\mathcal{M}\times\mathcal{M}$
kernel matrix is defined as 
\begin{equation}
\underline{\hat{\Lambda}}(\vec{\alpha})=\underline{G}^{-1/2}\left(\vec{\alpha}\right)\underline{\left\Vert \boldsymbol{\alpha}\right\rangle }\underline{\left\langle \boldsymbol{\beta}^{*}\right\Vert }\underline{G}^{-1/2}\left(\vec{\alpha}'\right),
\end{equation}
where $\underline{G}$ is a $\mathcal{M}\times\mathcal{M}$ diagonal
inner product matrix. 

From the above vector identities, one obtains 
\begin{align}
\hat{a}_{k}\underline{\hat{\Lambda}}(\vec{\alpha}) & =\underline{G}^{-1/2}\left(\vec{\alpha}\right)\alpha_{k}\underline{\sigma}_{k}\underline{\left\Vert \boldsymbol{\alpha}\right\rangle }\underline{\left\langle \boldsymbol{\beta}^{*}\right\Vert }\underline{G}^{-1/2}\left(\vec{\alpha}'\right)\nonumber \\
 & =\underline{\Sigma}\alpha_{k}\underline{\hat{\Lambda}}(\vec{\alpha}),
\end{align}
where 
\begin{equation}
\underline{\Sigma}=\underline{G}^{-1/2}\underline{\sigma}_{k}\underline{G}^{1/2},
\end{equation}
is the $\mathcal{M}\times\mathcal{M}$ renormalization matrix, and
we have used the identity $\underline{G}\underline{G}^{-1}=\underline{I}$. 

\subsection*{Anti-normally ordered kernel matrix identities:}

For a bosonic creation operator acting to the left of the kernel (anti-normal
ordering), one finds that: 
\begin{align}
\hat{a}_{k}^{\dagger}\hat{\Lambda}_{\tilde{p}\tilde{q}}\left(\overrightarrow{\alpha}\right) & =\bar{\sigma}_{k}^{\tilde{p}\tilde{p}'}\left(\partial_{k}\left\Vert \boldsymbol{\alpha}\right\rangle _{\tilde{p}'}\right)\left\langle \boldsymbol{\beta}^{*}\right\Vert _{\tilde{q}}e^{-w_{\tilde{p}\tilde{q}}\left(\overrightarrow{\alpha}\right)}\nonumber \\
 & =\bar{\Sigma}_{k}^{\tilde{p}\tilde{p}'}\left(\vec{\alpha}\right)\left(\partial_{k}+\frac{\partial w_{\tilde{p}'\tilde{q}}\left(\overrightarrow{\alpha}\right)}{\partial\alpha_{k}}\right)\hat{\Lambda}_{\tilde{p}'\tilde{q}}\left(\overrightarrow{\alpha}\right).
\end{align}
In this case the $\underline{\bar{\Sigma}}_{k}$ matrix increases
the $k$-th quantum numbers, and also includes the weight correction
term: 
\begin{equation}
\bar{\Sigma}_{k}^{\tilde{p}\tilde{p}'}\left(\vec{\alpha}\right)\equiv\bar{\sigma}_{k}^{\tilde{p}\tilde{p}'}e^{w_{\tilde{p}'}-w_{\tilde{p}}}.
\end{equation}

Similarly, for the hermitian conjugate case, one finds that if $\partial_{k}^{+}\equiv\partial/\partial\beta_{k}:$
\begin{align}
\hat{\Lambda}_{\tilde{p}\tilde{q}}\hat{a}_{k} & =e^{-w_{\tilde{p}\tilde{q}}}\left\Vert \boldsymbol{\alpha}\right\rangle _{\tilde{p}}\left(\frac{\partial}{\partial\beta_{k}}\left\langle \boldsymbol{\beta}^{*}\right\Vert _{\tilde{q}'}\right)\bar{\sigma}_{k}^{\tilde{q}'\tilde{q}}\nonumber \\
 & =\bar{\Sigma}_{k}^{\tilde{q}'\tilde{q}}\left(\vec{\alpha}'\right)\left(\partial_{k}^{+}+\frac{\partial w_{\tilde{p}\tilde{q}'}\left(\vec{\alpha}'\right)}{\partial\beta_{k}}\right)\hat{\Lambda}_{\tilde{p}\tilde{q}'}.
\end{align}

\subsubsection*{Weight derivative terms}

The above kernel matrix identities require weight derivatives. Following
the factorization Eq.(\ref{eq:weight_factorization}), these can be
written as follows;
\begin{align}
\frac{\partial w_{\tilde{p}\tilde{q}}}{\partial\alpha_{k}} & =\frac{1}{2}\frac{\partial\ln\left[G_{\tilde{p}}\left(\vec{\alpha}\right)G_{\tilde{q}}\left(\vec{\alpha}'\right)\right]}{\partial\alpha_{k}}\nonumber \\
 & =\frac{1}{2}\left[\frac{1}{G_{\tilde{p}}}\frac{\partial G_{\tilde{p}}}{\partial\alpha_{k}}+\frac{1}{G_{\tilde{q}}}\frac{\partial G'_{\tilde{q}}}{\partial\alpha_{k}}\right].
\end{align}
From the inner-product derivative result of Eq.(\ref{eq:inner-product-derivative}),
for the normal case with $G=G'$, this can be rewritten as:
\begin{align}
\frac{\partial w_{\tilde{p}\tilde{q}}}{\partial\alpha_{k}} & =\frac{\beta_{k}}{2}\left[\frac{G_{\tilde{p}-\tilde{h}_{k}}^{(N)}}{G_{\tilde{p}}^{(N)}}+\frac{G_{\tilde{q}-\tilde{h}_{k}}^{(N)}}{G_{\tilde{q}}^{(N)}}\right].\label{eq:normal-weight-derivative}\\
 & =\frac{\beta_{k}}{2}\left[T_{\tilde{p},k}+T_{\tilde{q},k}\right].
\end{align}

In the simple normalization case, the result is different, since $G\neq G'$:
\begin{align}
\frac{\partial w_{\tilde{p}\tilde{q}}}{\partial\alpha_{k}} & =\frac{\alpha_{k}^{*}}{2}\frac{G_{\tilde{p}-\tilde{h}_{k}}^{(S)}}{G_{\tilde{p}}^{(S)}}.\label{eq:simple-weight-derivative}\\
 & =\frac{\alpha_{k}^{*}}{2}T_{\tilde{p},k}^{(S)}
\end{align}

A matrix identity is obtained using these results as:
\begin{align}
\hat{a}_{k}^{\dagger}\underline{\hat{\Lambda}}(\vec{\alpha}) & =\underline{G}_{\tilde{p}}^{-1/2}\underline{\sigma}_{k}^{T}\left(\partial_{k}\underline{\left\Vert \boldsymbol{\alpha}\right\rangle }\right)\underline{\left\langle \boldsymbol{\beta}^{*}\right\Vert }\underline{G}_{\tilde{q}}^{-1/2}\nonumber \\
 & =\bar{\underline{\Sigma}}\left(\partial_{k}+\mathcal{T}_{k}\underline{\hat{\Lambda}}(\vec{\alpha})\right),
\end{align}
where in the ``normal'' case one needs a super matrix 
\begin{equation}
\mathcal{T}_{k}\underline{\hat{\Lambda}}(\vec{\alpha})\equiv\frac{1}{2}\left(\underline{T}_{k}\underline{\hat{\Lambda}}(\vec{\alpha})+\underline{\hat{\Lambda}}(\vec{\alpha})\underline{T}_{k}\right),
\end{equation}
with 
\begin{equation}
\bar{\underline{\Sigma}}=\underline{G}_{\tilde{p}}^{-1/2}\underline{\sigma}_{k}^{T}\underline{G}_{\tilde{p}}^{1/2},
\end{equation}
and for the ``simple'' normalization, one has:
\begin{equation}
\mathcal{T}_{k}\underline{\hat{\Lambda}}(\vec{\alpha})\equiv\frac{1}{2}\underline{T^{(S)}}_{k}\underline{\hat{\Lambda}}(\vec{\alpha}).
\end{equation}

\subsubsection*{Number conservation case}

As an example, let us treat number conservation arising from an $\mathcal{N}=1$
phase symmetry with $\mathcal{M\rightarrow\infty}$. 

In the normal weight case, we can simplify the identities through
defining an effective equivalent amplitude and photon number that
takes account of the projection index, as follows:
\begin{align}
\alpha_{p,j} & =\alpha_{j}\sqrt{p/n}\nonumber \\
\beta_{p,j} & =\beta_{j}\sqrt{p/n}\nonumber \\
n_{p,j} & =p\alpha_{j}\beta_{j}/n.
\end{align}

In terms of these new variables, the matrix-P elements and projectors
have identities for quadratic operators, as follows:
\begin{align}
\hat{a}_{j}\hat{\Lambda}_{pq}\hat{a}_{j}^{\dagger} & =\sqrt{n_{p,j}n_{q,j}}\hat{\Lambda}_{p-1,q-1}\nonumber \\
\hat{a}_{j}^{\dagger}\hat{a}_{j}\hat{\Lambda}_{pq} & =\left(\alpha_{j}\frac{\partial}{\partial\alpha_{j}}+\frac{1}{2}\left(n_{p,j}+n_{q,j}\right)\right)\hat{\Lambda}_{pq}\nonumber \\
\hat{\Lambda}_{pq}\hat{a}_{j}^{\dagger}\hat{a}_{j} & =\left(\beta_{j}\frac{\partial}{\partial\beta_{j}}+\frac{1}{2}\left(n_{p,j}+n_{q,j}\right)\right)\hat{\Lambda}_{pq},
\end{align}
where we recall the inner product definition $G_{p}^{(N)}=n^{p}/(p!)$. 

As an example, one can use these identities to evaluate the expectation
value of the photon number in channel $j$. One obtains, for S samples
labelled $(i)$, with a diagonal stochastic density matrix:
\begin{align}
\left\langle \hat{n}_{j}\right\rangle  & =\lim_{S\rightarrow\infty}\left\langle \hat{n}_{j}\right\rangle _{S}\nonumber \\
 & =\lim_{S,\mathcal{M}\rightarrow\infty}\frac{1}{S}\sum_{i=1}^{S}\sum_{p=1}^{\mathcal{M}}\Omega_{pp}^{(i)}n_{pj}^{(i)}.
\end{align}

\section*{Appendix D: Phase-space comparisons}

We now illustrate how the matrix-P method compares to previous methods
of representing bosonic quantum states using phase-space representations.
For comparison purposes, we now analyze how to sample the Schrödinger
cat state of Eq.(\ref{eq:Schrodinger-cat}) probabilistically, using
different phase-space methods that have positive distributions.

First consider three well-known techniques in classical phase-space,
for comparison purposes. The Glauber P-distribution is singular, the
Wigner distribution is non-positive and oscillatory, so neither can
be treated probabilistically. The Q-function is positive, with: 
\begin{align}
Q\left(\alpha\right) & =\frac{1}{\pi}\left|\left\langle \alpha\right.\left|\psi_{s}\right\rangle \right|^{2}\nonumber \\
 & =\frac{1}{2\pi C}e^{\left|\alpha\right|^{2}}\left|e^{\alpha x}+e^{-\alpha x}\right|^{2}.
\end{align}
This is a sum of Gaussians, which requires large sample numbers for
good accuracy.

\subsubsection{Nonclassical phase-space Schrödinger cat}

We now consider how to sample the sample the Schrödinger cat state
using different nonclassical phase-spaces, namely the positive-P and
matrix-P. The most compact positive P-distribution case has four terms,
each with two complex delta-functions \citep{Kiesewetter2017pulsed}:
\begin{equation}
P_{+}=\frac{e^{x^{2}}}{4C}\sum_{\pm}\delta(\alpha\pm x)\left[\delta(\beta\pm x)+\delta(\beta\mp x)e^{-2x^{2}}\right].
\end{equation}
This requires a summation over only four possible outputs, and is
more compact than the Q-function. 

The normal matrix-P distribution $P_{2}(\vec{\lambda},t)$ for this
case has a parity symmetry with $\mathcal{M}=2$, giving two parity
eigenvalues ($p=0,1$) and two projected subspaces consisting of number
states $\left|2n+p\right\rangle $ with either even or odd parity.
As a result, the projection operators are
\begin{equation}
\hat{\Pi}_{p}=\sum_{n=0}^{\infty}\left|2n+p\right\rangle \left\langle 2n+p\right|.
\end{equation}
On inspection of the power series expansion for the coherent states,
an un-normalized, projected coherent state is:
\begin{equation}
\left\Vert \alpha\right\rangle _{p}=\hat{\Pi}_{p}\left\Vert \alpha\right\rangle =\frac{1}{2}\left(\left\Vert \alpha\right\rangle +\left(-1\right)_{p}\left\Vert -\alpha\right\rangle \right).
\end{equation}
The kernel $\hat{\Lambda}_{pq}$ can be written in the coherent state
basis as:
\begin{equation}
\hat{\Lambda}_{pq}=\frac{e^{-w_{pq}}}{4}\left[\left\Vert \alpha\right\rangle +\left(-1\right)^{p}\left\Vert -\alpha\right\rangle \right]\left[\left\langle \beta\right\Vert +\left(-1\right)^{q}\left\langle -\beta\right\Vert \right].
\end{equation}
The most compact distribution is therefore a single delta-function
obtained by inspection of Eq.(\ref{eq:Matrix_representation})
\begin{equation}
P_{2}(\vec{\lambda},t)=\delta(\alpha-x)\delta(\beta-x)\delta(\underline{\Omega}-\tilde{\underline{\Omega}}),
\end{equation}
where $\tilde{\Omega}^{pq}=\delta_{p}\delta_{q}$ and $G_{0}^{(N)}=C$,
so $\text{Tr}_{2}\left[\tilde{\underline{\Omega}}\right]=1$ and $\int P_{2}(\vec{\lambda},t)\text{d}\vec{\lambda}=1.$

This mapping has a single term with no sampling errors. The result
from the existence theorem is not the optimal matrix-P distribution,
since it would have four terms, each duplicating the others.

\bibliographystyle{apsrev4-2}

\end{document}